\documentclass[useAMS,usenatbib]{mn2e}
\usepackage{graphicx}
\usepackage{bm}

\begin{document}
\title[Made-to-measure galaxy models II]{Made-to-measure galaxy models - II Elliptical and Lenticular Galaxies}
\author[R. J. Long and Shude Mao]
  {R. J.~Long$^{1,2}$\thanks{E-mail: rjl2007@gmail.com} and Shude~Mao$^{1,2}$\thanks{E-mail: smao@nao.cas.cn} \\
   $^1$National Astronomical Observatories, Chinese Academy of Sciences, A20 Datun Rd, Chaoyang District, Beijing 100012, China\\
   $^2$Jodrell Bank Centre for Astrophysics, Alan Turing Building, The University of Manchester, Manchester M13 9PL, UK\\}
\date{Accepted 2011 January 3.  Received 2011 November 7; in original form 2011 August 8}
\pagerange{\pageref{firstpage}--\pageref{lastpage}} \pubyear{2012}
\maketitle 

\label{firstpage}

\begin{abstract}
We take a sample of $24$ elliptical and lenticular galaxies previously analysed by the SAURON project using three-integral dynamical models created with Schwarzschild's method, and re-analyse them using the made-to-measure (M2M) method of dynamical modelling.  We obtain good agreement between the two methods in determining the dynamical mass-to-light (M/L) ratios for the galaxies with over $80\%$ of ratios differing by $< 10\%$ and over $95\%$ differing by $< 20\%$. We show that $\rmn{(M/L)_{M2M} \approx (M/L)_{Sch}}$.   For the global velocity dispersion anisotropy parameter $\delta$, we find similar values but with fewer of the made-to-measure models tangentially anisotropic by comparison with their SAURON  Schwarzschild counterparts. Our investigation is the largest comparative application of the made-to-measure method to date.
\end{abstract}


\section{Introduction}
Within the field of galactic and stellar dynamics, it has become common practice to model kinematic observations of a galaxy in order to interpret the observations and to understand better the underlying dynamical structures within the galaxy.  Within this modelling arena, the method of \citet{Schwarz1979} has been heavily developed and deployed with over $500$ citations from other papers (for example, \citealt{Rix1997}, \citealt{Bosch2008}, \citealt{Jalali2011}).  By comparison, the made-to-measure method (M2M) formulated by \citet{Syer1996} is less well-known but is no less capable, and has been the subject of growing interest recently (for example, \citealt{EJ2007}, \citealt{DL2007}, \citealt{DL2008a}, \citealt{Dehnen2009}, \citealt{Long2010}, \citealt{Das2011}).  Both methods achieve their objectives by weighting a system of particles / orbits and superimposing them to reproduce the galactic observations.  The key difference is that in Schwarzschild's method a library of orbits is first created and then weighted, whereas in the M2M method the orbit weights are determined dynamically as the particles are being orbited.  Other methods exist which, while not directly derived from \citet{Syer1996}, seek to tailor the kinematics of a system of particles to match the kinematics of a galaxy, for example \citet{Rodionov2009}.

In this paper we compare the made-to-measure method, as described by \citet{Long2010}, and Schwarzschild's method.  \citet{Cappellari2006} use Schwarzschild's method to determine the mass-to-light ratios for a selection of  galaxies observed with the SAURON \footnote{Spectrographic Areal Unit for Research on Optical Nebulae} integral-field spectrograph.  These same galaxies are re-analysed using the M2M method and the resulting ratios compared.  The galaxies comprise a mixture of elliptical and lenticular galaxies covering both fast and slow rotators, and including both edge-on galaxies and galaxies inclined to the line of sight.  As an extension to the mass-to-light exercise, we calculate the global anisotropy parameters as in \citet{SauronX2007} and again compare the results. Earlier papers (for example, \citealt{Das2011}, or \citealt{Lorenzi2009}) have used the M2M method effectively with individual galaxies.  To our knowledge, this paper is the first to use the M2M method with a larger sample of galaxies and is the first to compare directly the results achieved with those from using Schwarzschild's method.

In section \ref{sec:m2m} we describe the M2M method, and in section \ref{sec:sauronm2m} its application to the SAURON galaxies. Sections \ref{sec:ml} and \ref{sec:anisotropy} cover respectively the mass-to-light determinations and the global anisotropy parameters.  We draw the activities to a conclusion in section \ref{sec:conclusions}.  As might be expected, we refer heavily to the published SAURON material for data values.  We do not however cover in detail any theory from the SAURON material unless there is some specific point to be made in relation to the M2M method.  Unless otherwise stated we adopt the same modelling assumptions as \citet{Cappellari2006}.

\section{The M2M Method}\label{sec:m2m}

\subsection{Outline}
In brief, the M2M method is concerned with modelling stellar systems and individual galaxies as a system of test particles orbiting in a gravitational potential.  Weights are associated with the particles and are evolved over many orbital periods such that, by using these weights, observational measurements of a real galaxy are reproduced.  We expect that the weights themselves will have converged individually to some constant value.  It is natural to relate the particle weights to the luminosity of a galaxy and then to consider how the galaxy's surface brightness and luminosity weighted kinematics could be generated using the particle system.

In the next section, based on \citet{Long2010}, we set out the theory underlying the M2M method.

\subsection{Theory}
For a system of $N$ particles, orbiting in a gravitational potential, with weights $w_i$, the key equation which leads to the weight evolution equation is
\begin{equation}
	F(\bmath{w}) = -\frac{1}{2} \chi ^2 + \mu S + \frac{1}{\epsilon}\frac{dS}{dt} + \sum _i ^Q C_i
\end{equation}
where $\chi ^2$, $S$ and $C_i$ are all functions of the particle weights $\bmath{w} = (w_1, \cdot \cdot \cdot , w_N)$; $t$ is time; and $\mu$ and $\epsilon$ are positive parameters.
The equations governing weight evolution over time come from maximising $F(\bmath{w})$ with respect to the particle weights ($\partial F / \partial w_i = 0\;\; \forall i$) and rearranging terms to give equations of the form
\begin{equation}
	\frac{d}{dt} w_i = - \epsilon w_i G(\bmath{w}).
	\label{eqn:wtevoln}
\end{equation}
The overall rate of weight evolution is controlled by $\epsilon$.  The precise form of the function $G(\bmath{w})$ depends on the constraints $C_i$ and is illustrated later (equation \ref{eqn:gw}).  The process being applied to $\chi ^2$ is one of regularised, parameterised constrained extremisation.

The $\chi ^2$ term in $F$ arises from assuming that the probability of the model reproducing a single observation can be represented by a Gaussian distribution and then constructing a log likelihood function covering all observations. For $K$ multiple observables, we take $\chi ^2$ in the form
\begin{equation}
	\chi ^2 = \sum _k ^K \lambda _k \chi _k ^2
\label{eqn:chilambda}
\end{equation}
where $\lambda _k$ are small, positive parameters whose role is explained in section \ref{sec:params}.
\begin{equation}
	\chi ^2 _k = \sum _j ^{J_k} \Delta _{k,j} ^2
\end{equation}
and
\begin{equation}
	\Delta _{k,j} = \frac{y_{k,j}(\bmath{w}) -Y_{k,j}}{\sigma_{k,j}}
\end{equation}
where $Y_{k,j}$ is the measured value of observable $k$ at position $j$ with error $\sigma_{k,j}$, and $y_{k,j}(\bmath{w})$ is the model equivalent of $Y_{k,j}$.
\begin{equation}
	y_{k,j}(\bmath{w}) = \sum _i ^N w_i K_{k,j}(\bmath{r}_i, \bmath{v}_i) \delta(i \in k,j)
\end{equation}
where $K_{k,j}(\bmath{r}_i, \bmath{v}_i)$ is the kernel for observable $k$ evaluated at position $j$ for a particle with position $\bmath{r}_i$ and velocity $\bmath{v}_i$.  $\delta(i \in k,j)$ is a selection function and signifies that only particles which contribute to observable $k$ at position $j$ should be included in the calculation of $y_{k,j}$.  We have listed the kernels required for this paper in section \ref{sec:kernels}.

The entropy function $S$ in $F$ is
\begin{equation}
	S(\bmath{w}) = - \sum _i ^N w_i \ln (\frac{w_i}{m_i})
\end{equation}
where $m_i$ is taken as the initial value of a particle weight (in practice, we take $m_i = 1/N$).  $S$ is used for regularisation / smoothing purposes with the amount of regularisation being controlled by the parameter $\mu$. The derivative term $dS/dt$ indicates that over time we require the particle weights, and thus $S$, to be constant.  As demonstrated in \citet{Long2010}, the term behaves as the constraint $dS/dt = 0$.

The functions $C_i$ in $F$ are additional constraints to be included in the maximisation of $F$.  In this paper, we use only one such constraint which is that we require the model luminosity to match the luminosity ($L$) of the galaxy being modelled, that is $\sum Lw_i = L$, or more concisely $\sum w_i = 1$.  We therefore take
\begin{equation}
	C_1 = - \frac{\lambda_{\rmn{sum}}}{2}  \left ( \sum _i ^N  w_i - 1 \right ) ^2
\end{equation}
where $\lambda_{\rmn{sum}}$ is a positive parameter.

Given the definitions of $\chi ^2$, $S$ and $C_i$ and noting that for the purposes of this paper we do not use regularisation ($\mu = 0$), $G(\bmath{w})$ from equation \ref{eqn:wtevoln} can now be written
\begin{equation}
	G(\bmath{w}) = \sum _k ^K \lambda _k \frac{K_{k,ji}}{\sigma _{k,j}} \Delta _{k,j} + \lambda_{\rmn{sum}} \left ( \sum _ i ^N w_i - 1 \right )
\label{eqn:gw}
\end{equation}
where $K_{k,j}(\bmath{r}_i, \bmath{v}_i)$ has been abbreviated to $K_{k,ji}$ and we have assumed that, for all observables, 1 particle contributes only at 1 position $j$.

Finally, model observables (and thus particle weights) are subject to noise as the numbers of particles contributing to the observables vary.  This noise is suppressed by replacing $\Delta _{k,j}$ in $G(\bmath{w})$ by an exponentially smoothed version $\tilde{\Delta}_{k,j}$ given by
\begin{equation}
  \frac{d}{dt} \tilde{\Delta}_{k,j} = \alpha ( \Delta _{k,j} - \tilde{\Delta}_{k,j} )
\end{equation}
where $\alpha$ is a small positive parameter.  The smoothed $\Delta _{k,j}$ can be used to calculated a smoothed version $\tilde{y} _{k,j}$ of the model observable,
\begin{equation}
	\tilde{y} _{k,j} = Y_{k,j} + \sigma _{k,j} \tilde{\Delta}_{k,j}.
\end{equation}

\section{SAURON M2M Models}\label{sec:sauronm2m}

\subsection{Galaxies and observables}\label{sec:galaxies}
The galaxies which we model are as in \citet{Cappellari2006} but with NGC 221 omitted since it is not part of the SAURON data release.  The galaxies are listed in Table \ref{tab:galaxysample} together with the properties which are relevant to M2M modelling.  As indicated earlier, the galaxies comprise a mixture of elliptical and lenticular galaxies covering both fast and slow rotators, and including both edge-on galaxies and galaxies inclined to the line of sight.  The galaxies also exhibit various core features, for example kinematically distinct cores or counter rotating cores \citep{Emsellem2004}.

The inclinations and distances (distance modulus) to the galaxies are as per \citet{Cappellari2006} Table 1. We have not attempted to use M2M modelling to determine the inclinations.  Within a M2M model, we employ Cartesian axes such that the positive x-axis points towards the observer and the $y-z$ plane represents the galaxy's on sky projection.  We align the galaxies' photometric major axes to the model y-axis utilising position angles taken from \citet{SauronX2007}.

We take kinematic data for the galaxies from the SAURON data release \citep{Emsellem2004}. The data available are the line-of-sight mean velocity, velocity dispersion and the $h_3$ and $h_4$ Gauss-Hermite coefficients, all taken from a truncated Gauss-Hermite expansion of the line-of-sight velocity distribution \citep{Vandermarel1993},
\begin{equation}
	\rmn{losvd}(v) =  \frac{\exp(- v_{\rmn{norm}}^2 / 2)}{\sigma \sqrt{2 \pi}} \left [ 1 + \sum _{n=3} ^{4} h_n H_n(v_{\rmn{norm}}) \right ]
\end{equation}
where $H_n$ is the Hermite polynomial of degree $n$ and the normalised velocity $v_{\rmn{norm}}$ is defined as
\begin{equation}
	v_{\rmn{norm}} = \frac{v - \bar{v}}{\sigma}
\end{equation}
where $\bar{v}$ and $\sigma$ are the line-of-sight mean velocity and velocity dispersion respectively.

The $h_5$ and $h_6$ Gauss-Hermite coefficients used in \citet{Cappellari2006} are not available in the data release. We assume therefore that $h_5 = h_6 = 0$ with a measurement error of $\pm 0.3$ (M. Capellari private communication).  We do not model a galaxy's mean line-of-sight velocity and velocity dispersion directly but instead model $h_1 = h_2 = 0$ as in \citet{Rix1997}.  Following \citet{Magorrian1994}, we calculate the measurement errors $\Delta h_1$ and $\Delta h_2$ as
\begin{equation}
	\Delta h_1 = - \frac{\Delta v}{\sqrt{2} \sigma}
\end{equation}
and
\begin{equation}
	\Delta h_2 = \frac{\Delta \sigma}{2 \sigma} \left ( \sqrt{12} h_4 - \sqrt{2} \right ),
\end{equation}
where $\Delta v$ and $\Delta \sigma$ respectively are the measurement errors in the mean line-of-sight velocity $v$ and velocity dispersion $\sigma$.  If we require the model mean line-of-sight velocity $v_m$ or the model line-of-sight velocity dispersion $\sigma _m$, we calculate them as
\begin{equation}
	v_m = v - \sqrt{2} \sigma h_{1,m}
	\label{eqn:modelv}
\end{equation}
and
\begin{equation}
	\sigma _m = \sigma + \frac{2 \sigma h_{2,m}}{\sqrt{12} h_{4,m} - \sqrt{2}},
	\label{eqn:modelsigma}
\end{equation}
where the $h_{i,m}$ are the exponentially smoothed model $h_i$ values.  \citet{Lorenzi2009} use a similar approach in their M2M models of NGC 3379.  It is possible to calculate $v_m$, $\sigma _m$ and the Gauss-Hermite coefficients by fitting Gauss-Hermite series directly to the end of modelling run particle data, provided sufficient particles are available to populate the velocity histograms necessary to the fitting process.  The approach above, using smoothed model values, avoids the need to run the M2M models with large numbers of particles.

We put the SAURON kinematic data through a cleaning process (section \ref{sec:dataclean}) before subtracting the systemic galactic velocity from the mean line-of-sight velocity, symmetrizing the data and converting it to units appropriate to our M2M modelling (distances in effective radii, time in $10^7$ years).  We take the systemic velocities from \citet{Emsellem2004} and assume they are subject to a measurement error of $10\%$. The usual error propagation rules are applied.

The observables in our M2M models are thus
\begin{enumerate}
\item surface brightness,
\item Gauss-Hermite coefficients $h_1$ to $h_6$.
\end{enumerate}
Values for the kinematic observables are as described above, and surface brightness is calculated from the multi-Gaussian expansions of the galaxy's surface brightness (see section \ref{sec:potentials}).  For modelling purposes, we assume a $10\%$ relative error in surface brightness values. Unless explicitly stated, luminosity density is not used in our M2M models to constrain the luminous matter distribution (see section \ref{sec:misc}).

Similarly to \citet{Cappellari2006}, we perturb the line-of-sight particle coordinates by a `point spread function' before binning any model data to create the model observables.  We use the seeing values from \citet{Emsellem2004} Table 3 and implement the function as a circular Gaussian distribution.

\begin{table*}
	\centering
	\caption{Galaxy sample and measured parameters}
	\label{tab:galaxysample}
	\begin{tabular}{llcccccccccc}
		\hline
Galaxy & Type & Fast & Distance & $R_e$  & $\sigma _e$ & $i$ & $\rmn{PA}$ & Seeing & $V_{\rmn{syst}}$ & $(M/L)_{\rmn{Schw}}$ & $(M/L)_{\rmn{M2M}}$\\
   & & rotator & Mpc & arcsec & $\rmn{km\ s^{-1}}$ & deg & deg & arcsec & $\rmn{km\ s^{-1}}$ & I-band & I-band \\
(1) & (2) & (3) & (4) & (5) & (6) & (7) & (8) & (9) & (10) & (11) &(12)\\
		\hline
NGC 524  & $\rmn{S0^+(s)}$ & yes & 23.34 & 51 & 235 & 19 & 48.4 & 1.4 & 2353 & 4.99 & 6.39    \\
NGC 821  & $\rmn{E6?}$ & yes & 23.44 & 39 & 189 & 90 & 32.2 & 1.7 & 1722 & 3.08 & 3.37  \\
NGC 2974 & $\rmn{E4}$ & yes & 20.89 & 24 & 233 & 57 & 43.5 & 1.4 & 1886 & 4.52 & 4.60  \\
NGC 3156 & $\rmn{S0:}$ & yes & 21.78 & 25 & 65 & 67 & 49.4 & 1.6 & 1541 & 1.58 & 1.46   \\
NGC 3377 & $\rmn{E5-6}$ & yes & 10.91 & 38 & 138 & 90 & 41.3 & 2.1 & 690 & 2.22 & 2.22   \\
NGC 3379 & $\rmn{E1}$ & yes & 10.28 & 42 & 201 & 90 & 67.9 & 1.8 & 916 & 3.36 & 3.67   \\
NGC 3414 & $\rmn{S0 pec}$ & no & 24.55 & 33 & 205 & 90 & 179.9 & 1.4 & 1472 & 4.26 & 4.56\\
NGC 3608 & $\rmn{E2}$ & no & 22.28 & 41 & 178 & 90 & 79.3 & 1.5 & 1228 & 3.71 & 3.73   \\
NGC 4150 & $\rmn{S0^0(r)?}$ & yes & 13.37 & 15 & 77 & 52 & 147.0 & 2.1 & 219 & 1.30 & 1.26  \\
NGC 4278 & $\rmn{E1-2}$ & yes & 15.63 & 32 & 231 & 45 & 16.7 & 1.9 & 631 & 5.24 & 5.61   \\
NGC 4374 & $\rmn{E1}$ & no & 17.87 & 71 & 278 & 90 & 128.2 & 2.2 & 1023 & 4.36 & 4.65   \\
NGC 4458 & $\rmn{E0-1}$ & no & 16.75 & 27 & 85 & 90 & 4.5 & 1.6 & 683 & 2.28 & 2.32\\
NGC 4459 & $\rmn{S0^+(r)}$ & yes & 15.70 & 38 & 168 & 47 & 102.7 & 1.5 & 1200 & 2.51 & 2.76  \\
NGC 4473 & $\rmn{E5}$ & yes & 15.28 & 27 & 192 & 73 & 93.7 & 1.9 & 2249 & 2.91 & 3.12   \\
NGC 4486 & $\rmn{E0-1^+ pec}$ & no & 15.63 & 105 & 298 & 90 & 158.2 & 1.0 & 1274 & 6.10 & 7.05   \\
NGC 4526 & $\rmn{SAB0^0(s)}$ & yes & 16.44 & 40 & 222 & 79 & 112.8 & 2.8 & 626 & 3.35 & 3.26 \\
NGC 4550 & $\rmn{SB0^0:sp}$ & yes & 15.42 & 14 & 110 & 84 & 178.3 & 2.1 & 413 & 2.62 & 2.78  \\
NGC 4552 & $\rmn{E0-1}$ & no & 14.93 & 32 & 252 & 90 & 125.3 & 1.9 & 351 & 4.74 & 5.01   \\
NGC 4621 & $\rmn{E5}$ & yes & 17.78 & 46 & 211 & 90 & 163.3 & 1.6 & 456 & 3.03 & 3.07  \\
NGC 4660 & $\rmn{E}$ & yes & 12.47 & 11 & 185 & 70 & 96.8 & 1.6 & 1089 & 3.63 & 3.85  \\
NGC 5813 & $\rmn{E1-2}$ & no & 31.33 & 52 & 230 & 90 & 134.5 & 1.7 & 1947 & 4.81 & 4.69 \\
NGC 5845 & $\rmn{E:}$ & yes & 25.24 & 4.6 & 239 & 90 & 143.2 & 1.5 & 1474 & 3.72 & 4.34 \\
NGC 5846 & $\rmn{E0-1}$ & no & 24.21 & 81 & 238 & 90 & 75.2 & 1.4 & 1710 & 5.30 & 5.38  \\
NGC 7457 & $\rmn{S0^-(rs)?}$ & yes & 12.88 & 65 & 78 & 64 & 125.5 & 1.3 & 845 & 1.78 & 1.52  \\
		\hline
	\end{tabular}
	
\medskip
\flushleft
Column (1): NGC number.
Column (2): Hubble type from \citet{Emsellem2004}.
Column (3): Fast/slow rotator from \citet{Cappellari2006}.
Column (4): Distance calculated from distance modulus in \citet{Cappellari2006}.
Column (5): Effective (half-light) radius from \citet{Cappellari2006}.
Column (6): Velocity dispersion within effective radius from \citet{Cappellari2006}.
Column (7): Inclination from \citet{Cappellari2006}.
Column (8): Position angle from \citet{SauronX2007}.
Column (9): Seeing from \citet{Emsellem2004}.
Column (10): Systemic velocity from \citet{Emsellem2004}.
Column (11): Schwarzschild mass-to-light ratio from \citet{Cappellari2006}.
Column (12): M2M mass-to-light ratio - section \ref{sec:ml}.
\end{table*}

\subsection{Voronoi Tessellation}
To achieve a pre-determined signal to noise, the SAURON observations were adaptively binned and processed, as described in \citet{Cappellari2003} and \citet{Emsellem2004}, resulting in a centroidal Voronoi tessellation. The kinematic data in the SAURON data release are presented in the context of that tessellation. The M2M models use the same Voronoi tessellation for determining model kinematic observables. The Voronoi cells are the bins used for accumulating particle kinematic data as part of the construction of the model observables.

\begin{figure}
		\centering
		\includegraphics[width=84mm]{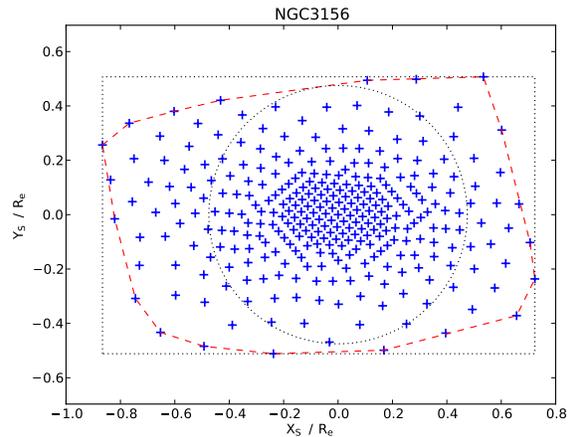}
		\caption[Kinematic measurement points for NGC 3156]{Measurement points $(X_S, Y_S)$ are in blue and the convex hull is indicated by the red dashed lines.  The bounding rectangle is used in calculating the bin areas.  The circle has the maximum possible radius for a circle to lie within the convex hull, and is used for performance reasons as part of the algorithm to determine whether or not a particle is inside the hull or not. Distances are in units of effective radii $R_e$ for the galaxy.}
	\label{fig:hull}
\end{figure}

Within our M2M implementation, the Voronoi bins are represented by a 2D `tree' with a surrounding convex hull.  Particles are binned by first determining whether they are inside the convex hull and, if they are, performing a `nearest neighbour search' of the 2D tree to identify the bin required.  The areas of the bins (needed for model observable calculations - see section \ref{sec:kernels}) are calculated using a Monte Carlo approach. As an example, Figure \ref{fig:hull} shows the positions of the Voronoi centroids (the data measurement points) for NGC 3156 together with their convex hull.  The SAURON data release contains no information on the extent of the outermost Voronoi bins, and, as a consequence, the model observable calculations for such bins are biased inwards.  This is not considered a significant issue as the number of such bins is small.

Voronoi bins are not used with surface brightness.  Instead we employ a $16 \times 16$ polar grid with radial divisions pseudo-logarithmic as described \citet{Long2010}.

To give an indication of the number of observations being modelled, of the 24 galaxies, NGC 5845 has the least number of Voronoi bins with $164$ and NGC 4486 the greatest with $2112$ bins.  The total number of observations is therefore $1240$ for NGC 5845 and $12,928$ for NGC 4486.

\subsection{Data cleaning}\label{sec:dataclean}
We have subjected the SAURON data release to a number of checks to eliminate suspect kinematic data
\begin{enumerate}
\item `nan' (not a number) value check
\item zero value check, for example error fields should not be zero
\item positive value check, for example error fields should not be negative
\item small value check ($< 10^{-5}$)
\item record sequence check - a record is flagged if a data field in one record has the same value as in the previous record
\end{enumerate}
Note that not all checks are applied to all data fields.

In total, $13$ galaxies were found to have suspect data as a result of the exercise, and consequently we have not used the associated Voronoi data bins in the modelling process.  The bins are not actually removed from the Voronoi tessellations but marked as `not in use'.

As identified in \citet{Emsellem2004}, the NGC 5846 data contain contamination from a foreground star and a companion (NGC 5846A).  This north and south contamination has, as far as possible, been removed by simply deleting all the Voronoi bins where the magnitude of the y-coordinate position is greater than $3.5$ kpc.

\subsection{Gravitational Potentials}\label{sec:potentials}
All the galaxies are modelled as axisymmetric galaxies with their gravitational potentials calculated from deprojecting the multi-Gaussian expansions (MGEs) of their surface brightness recorded in \citet{Cappellari2006}, and \citet{Krajnovic2005} for NGC 2974.  The multi-Gaussian expansion technique is described in \citet{Emsellem1994} and is not repeated here.  The galactic kinematic and photometric symmetry axes are assumed to align.  No M2M modelling of the galaxies with the axes not aligned, as discussed in \citet{SauronIX2007} for example, is undertaken.

In our M2M implementation, to avoid multiple numerical integrations, the MGE potential and associated accelerations are pre-calculated and held on $1500 \times 1500$ $R - z$ interpolation grids with bilinear interpolation used between grid points. We employ OpenMP \footnote{http://openmp.org} to accelerate production of the grids.

We augment the MGE potential with a central black hole modelled as a Keplerian potential. The mass of the black hole $M_{BH}$ is calculated using the $M_{BH}$ - $\sigma$ relationship as described in \citet{Gultekin2009} ($\sigma$ is the bulge velocity dispersion).  For consistency with \citet{Cappellari2006} we do not include a dark matter component in the potential.

Orbit integration is performed using the standard $2^{nd}$ order interleaved leap frog method with an adaptive time step.  The duration of our M2M models is inversely proportional to the dynamical time of the galaxy being modelled and is numerically, approximately $400$ divided by the dynamical time with a minimum of 200 units.  We use the formula for dynamical time in \citet{BT2008} \textsection 2.2.2 and calculate it at the half light radius of the model.  The size of a galaxy model is $3$ times the maximum dispersion in the galaxy's surface brightness multi-Gaussian expansion.  In practice, the model sizes range from $5$ to $18$ effective radii depending on the galaxy.

\subsection{Particle initial conditions}\label{sec:particleics}
In setting the initial spatial and velocity coordinates for particles, two issues need to be addressed.  The first is how to handle global rotation of the galaxy, and the second, how to handle core features.  In both cases, we may choose to take no explicit action and allow the M2M method to attempt to weight the particles such that the observables are reproduced.  Alternatively, we may use our knowledge of the galaxy's features and set the initial particle conditions accordingly.  The first approach, taking no explicit action, is inefficient in the use of particles (consider the case of many particles orbiting in the opposite sense to any global rotation - the method will lower the particles' weights to reduce the particles' influence on the model observables).  We therefore discount the first approach and adopt the second.

\begin{figure*}
		\centering
		\includegraphics[width=84mm]{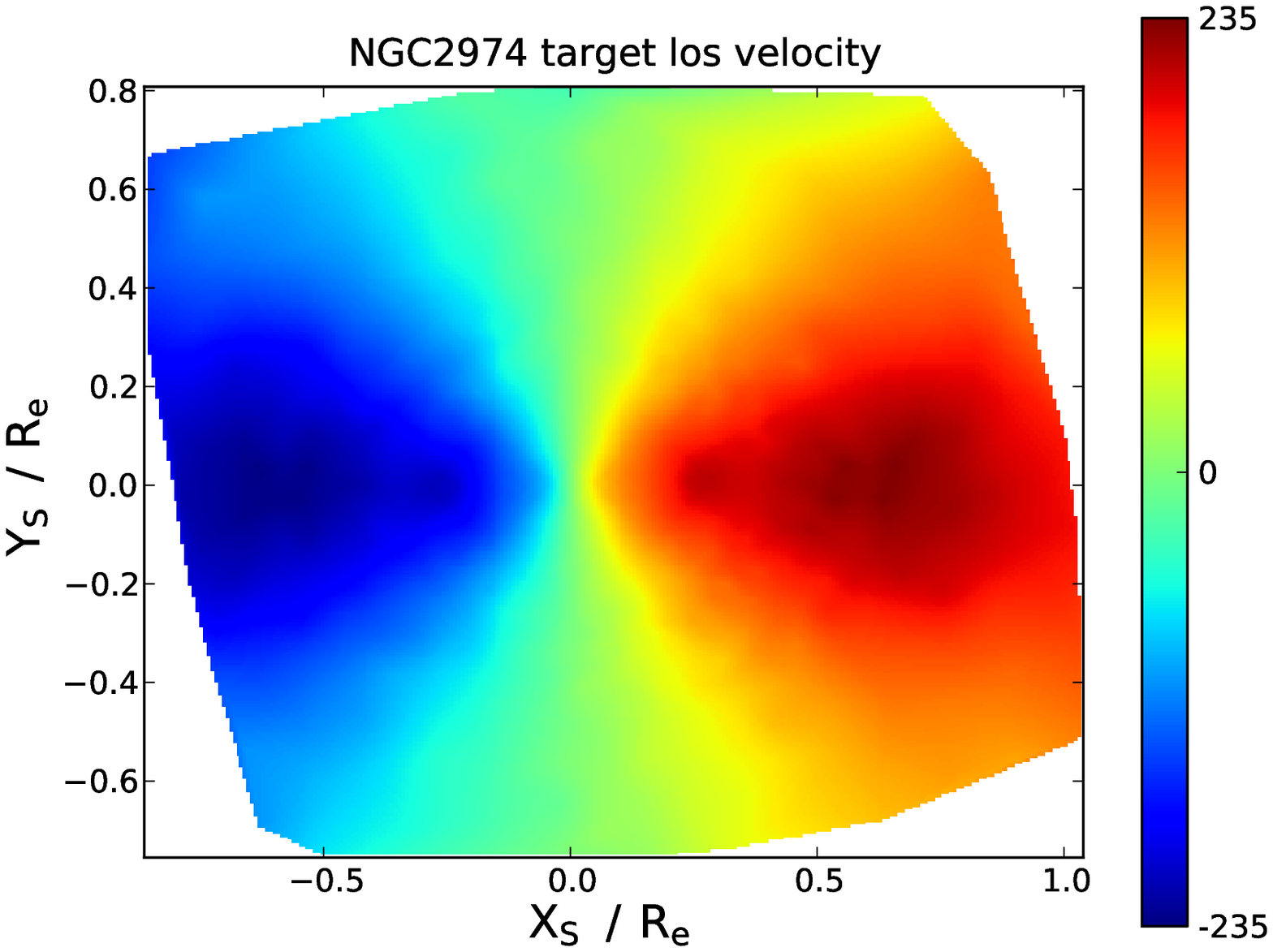}
		\includegraphics[width=84mm]{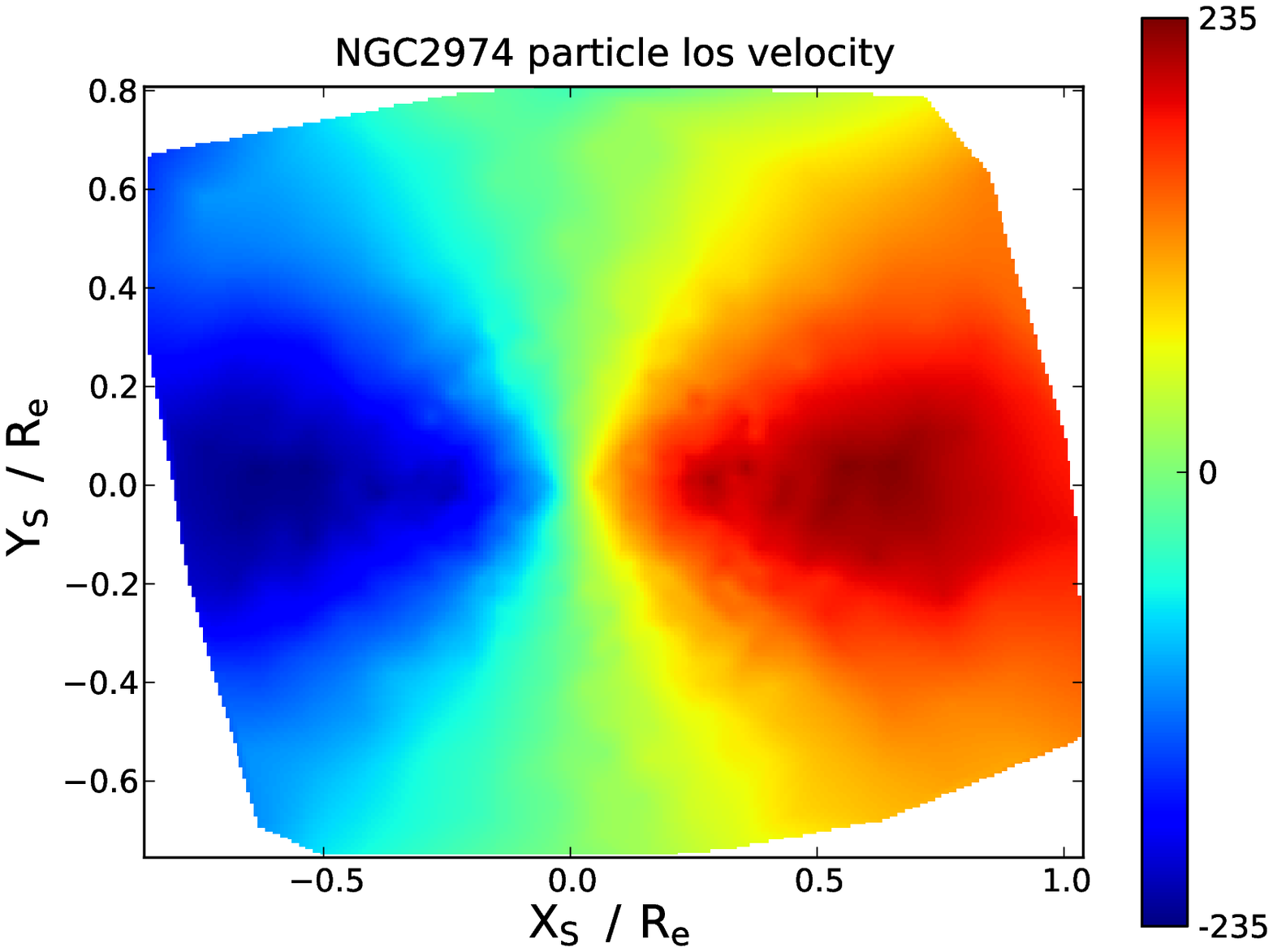}
		\caption[Particle initial conditions]{For NGC 2974, the left panel shows the symmetrised measured mean line-of-sight velocity and the right panel, the line-of-sight velocity from the particle initial conditions (at the start of modelling). The velocity units are $\rmn{km\ s^{-1}}$.}
\label{fig:icsNGC2974}
\end{figure*}

We employ two schemes for setting the initial conditions.  In the first, we set the initial spatial positions of the particles to approximate the luminosity distribution generated by deprojecting the galaxy's surface brightness MGE.  Creation of the velocity coordinates follows a 3 stage process,
\begin{enumerate}
\item use the velocity dispersions created by solving the semi-isotropic Jeans equations for an axisymmetric system \citep{BT2008} to provide initial values for the velocity coordinates,
\item set the global rotation sense for a prespecified fraction of the particles to align with the rotation sense of the SAURON observations, and
\item for the particles inside the SAURON measurements convex hull, adjust the coordinates to approximate the measured line-of-sight velocity by sampling from a Gaussian distribution formed from the measured line-of-sight velocity and dispersion.
\end{enumerate}
The effect of stage (iii) is to reproduce (approximately) in the particle initial conditions any core features in the SAURON velocity measurements.  Determination of the `prespecified fraction' in stage (ii) is not yet an automated process.  The fraction is determined iteratively by comparing visually the SAURON velocity contours with the particle equivalents and then adjusting the fraction as necessary.  As an example, we show the resulting velocity contours for NGC 2974 in Figure \ref{fig:icsNGC2974}.  

For the second scheme, a grid-less energy, angular momentum and pseudo third integral system similar to that in  \citet{Cappellari2006} is adopted to determine the initial conditions for the particle orbits. We then modify the initial conditions as in stages (ii) and (iii) above.  Unless otherwise stated, by default, all modelling runs are performed using this scheme.

We determine the number of particles to be used in our M2M models by examining the particle distribution across the Voronoi bins.  We adjust the number of particles such that the minimum mean number of particles per bin is greater than $60$. We find that between $10^5$ and $5 \times 10^5$ particles are required depending on the galaxy. Ideally the minimum number of particles per bin should be used but this would require much larger numbers of particles ($>10^6$) and correspondingly more computing resources.  Note that at any one time during a modelling run significant numbers of particles ($> 30\%$) are outside of the convex hull and are not contributing to reproducing the kinematic observables.

\subsection{Kernels}\label{sec:kernels}
The kernels are similar to those in \citet{Long2010} and we list them below.
\begin{enumerate}
\item luminosity density
\begin{equation}
	K_{ji} = \frac{L}{V_j}
\end{equation}

\item surface brightness
\begin{equation}
	K_{ji} = \frac{L}{A_j}
\end{equation}

\item mean luminosity-weighted Gauss-Hermite coefficient $h_n$
\begin{equation}
	K_{ji} = \frac{\sqrt 2  L}{I_j A_j} H_n(v_{\rmn{norm},ji}) \exp(- v_{\rmn{norm},ji}^2 / 2)
\end{equation}
\end{enumerate}
where $L$ is the luminosity of the galaxy being modelled, $A_j$ is the area of the bin at position $j$ and $I_j$ the target surface brightness, $V_j$ is the bin volume, $v_{\parallel, i}$ is the line-of-sight velocity for particle $i$, and $H_n$ is the Hermite polynomial of degree $n$.  The normalised velocity $v_{\rmn{norm},ji}$ is defined as
\begin{equation}
	v_{\rmn{norm},ji} = \frac{v_{\parallel, i} - \bar{v}_j}{\sigma _j}
\end{equation}
where $\bar{v}_j$ and $\sigma _j$ are the measured line-of-sight mean velocity and velocity dispersion respectively.  We normalise the Hermite polynomials as described in \citet{Vandermarel1993} appendix A.

\subsection{Parameter setting}\label{sec:params}
In this section we describe how we set the values of the various parameters within the M2M method.  As indicated earlier, we do not use regularisation for any of the modelling runs and so we take $\mu = 0$. For exponential smoothing we take a common approach across all the galaxies and set $\alpha = 0.05$.  The $\epsilon$ parameter controls the overall rate of weight evolution and we set $\epsilon = 10^{-4}$ initially.  We may alter it later if we find that we are not achieving a $\chi ^2 _k$ per degree of freedom value of $O(1)$ for a modelling run.  Finally, we set $\lambda_{\rmn{sum}} = 10^4$.

The role of the observable $\lambda _k$ parameters (see equations \ref{eqn:chilambda} and \ref{eqn:gw}) is to help balance the weight evolution equation across all the observables.  The equation contains terms of the form
\begin{equation}
	\lambda _k  \frac{K_{k,ji}}{\sigma _{k,j}} \Delta _{k,j}.
\end{equation}
The $\Delta _{k,j}$ component reflects how well the model is reproducing the measured observations and is not examined further.  The $K_{k,ji} / \sigma _{k,j}$ component varies, by several orders of magnitude, between observables and between positions $j$ for a single observable.  By running a M2M model for a short period of time ($5$ dynamical time units), we are able to understand how the $K_{k,ji} / \sigma _{k,j}$ values are varying.  We take the modal value of $K_{k,ji} / \sigma _{k,j}$ (found by binning logarithmically) and set $\lambda _k$ such that
\begin{equation}
	\lambda _k  \frac{K_{k,ji}}{\sigma _{k,j}} |_{\rmn{modal}} = 10.
\end{equation}
Similarly to $\epsilon$, we may adjust the value of $\lambda _k$ if we find that we are not achieving a $\chi ^2 _k$ per degree of freedom value of $O(1)$ for a modelling run. 

For the $24$ galaxies we analysed, involving some $168$ $\lambda _k$'s, approximately $25\%$ of the $\lambda _k$'s required adjustment.  \citet{Cappellari2006} noted that reproducing the Gauss-Hermite coefficient $h_4$ proved problematic. Based on the $\chi _{h_4} ^2$ values we achieve, we do not have an equivalent issue with $h_4$.

\subsection{Computer performance}
Our M2M software has been parallelised using the Message Passing Interface (MPI) with the parallelisation being based around a star network with a single central controlling node.  We reported in \citet{Long2010} that the implementation was highly scaleable and others, for example \citet{Dehnen2009} and \citet{DL2007}, have reported similarly.  This position remains true for a low number of observables and measurement points.  However given the number of measurements available for the galaxies we have analysed (see section \ref{sec:galaxies}), we find that the scaleability is reduced.  This reduction in our case is due to the overheads of handling data packet fragmentation particularly on the central node of the network.  Increasing the packet size will recover some of the reduction.  For larger M2M models, it may be appropriate to introduce a layer of nodes whose primary role is to act as data concentrators.  We have yet to investigate either of these schemes.

\subsection{Miscellany}\label{sec:misc}
As indicated earlier, this paper builds on \citet{Long2010} (Paper 1).  In this section, we identify various mechanisms and results from that paper, relevant to the current investigation, which have not been dealt with elsewhere.

We use the same weight convergence assessment mechanism as in Paper 1.  It is important to note that if a M2M model reproduces the constraining observables, this does not necessarily mean that the particle weights have converged.

We demonstrated in Paper 1 that using regularisation ($\mu \neq 0$) and luminosity density as a constraint made little difference to the model determined mass-to-light ratio.  We have chosen not to use regularisation in this paper.  \citet{Cappellari2006} did in fact use integral space second derivative regularisation in their mass-to-light exercise.  Similarly to regularisation, we choose not to use luminosity density as a constraint on the luminous matter distribution within our M2M models as a matter of course, but include it, where explicitly stated, for comparison purposes only.

For consistency with \citet{Cappellari2006}, we quote no confidence intervals on our model determined mass-to-light ratios.

\section{Mass-to-light Determination}\label{sec:ml}
The process for determining the mass-to-light ratio for a galaxy is straightforward and widely used elsewhere.  We run a series of M2M models varying a mass-to-light parameter ($\Upsilon$) and look for a minimum in the resulting model $\chi ^2$ values.  The parameter value at the minimum, adjusted for the black hole mass, is taken as the `true' mass-to-light ratio for the galaxy given all the modelling assumptions.
\begin{equation}
	\Upsilon _{\rm{adjusted}} = \frac{\Upsilon _{\rm{model}} L + M_{\rm{BH}}}{L}
\end{equation}
where $L$ is the model total luminosity and $M_{\rm{BH}}$ is the black hole mass.

\begin{figure}
		\centering
		\includegraphics[width=84mm]{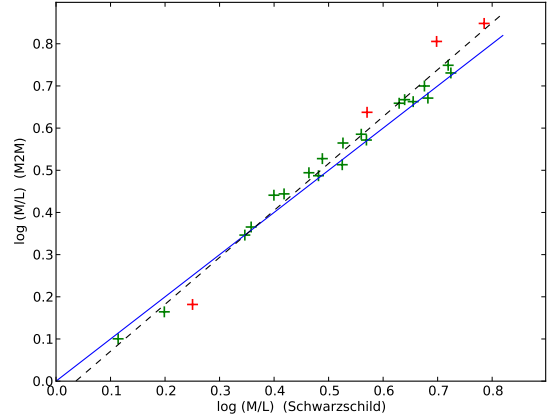}
		\caption[Comparison of M2M vs Schwarzschild mass-to-light ratios]{Comparison of M2M vs Schwarzschild mass-to-light ratios with galaxies which differ by more than $10\%$ marked in red. The solid blue line represents equality of the ratios, and the black dashed line a least squares fit to the data.}
	\label{fig:mlplot15}
\end{figure}

The mass-to-light ratios we achieve are shown in Table \ref{tab:galaxysample}, and Figure \ref{fig:mlplot15} contains a plot of the M2M mass-to-light values against those achieved by \citet{Cappellari2006} using Schwarzschild's method. For $96\%$ of the galaxies, the relative difference between the mass-to-light values is $< 20\%$ and for $83\%$ of the galaxies, the difference is  $< 10\%$.  $3$ galaxies have differences $> 15\%$ - NGC 524 ($28\%$), NGC 4486 ($16\%$), and NGC5845 ($17\%$). Table \ref{tab:reldiff} contains a fuller breakdown.

\begin{table}
	\centering
	\caption{Comparison between M2M and SAURON mass-to-light ratios}
	\label{tab:reldiff}
	\begin{tabular}{cccl}
		\hline
		Difference & Number of &  & Galaxy\\
		           & galaxies  &  &  NGC numbers\\
		\hline
		$< 5\%$ & $9$ & $38\%$ & 2974, 3377, 3608, 4150, \\
		& & & 4458, 4526, 4621, 5813, \\
		& & & 5846\\
		$\;\,5\%$ to $10\%$ & $11$ & $46\%$ & 821, 3156, 3379, 3414,\\
		& & & 4278, 4374, 4459, 4473, \\
		& & & 4550, 4552, 4660\\
		$10\%$ to $15\%$ &  $1$ & $4\%$ & 7457\\
		$> 15\%$ & $3$ & $13\%$ & 524, 4486, 5845\\
		\hline
	\end{tabular}
	
\medskip
\end{table}

Performing a least squares straight line fit to the logarithmic mass-to-light data yields
\begin{equation}
	\Upsilon _{\rmn{M2M}} \propto \Upsilon _{\rmn{Sch}} ^{1.11 \pm 0.04}
\end{equation}
where $\Upsilon _{\rmn{M2M}}$ is the M2M mass-to-light ratio and $\Upsilon _{\rmn{Sch}}$ is the Schwarzschild equivalent. Introducing luminosity density to constrain the distribution of luminous matter does not alter this relationship. (For most galaxies, the models already well reproduce the density without the use of an explicit constraint).  Changing the particle initial conditions to scheme 1 in section \ref{sec:particleics} (match the luminosity density spatially, velocities based on the Jeans' equations) does not significantly alter the relationship with
\begin{equation}
	\Upsilon _{\rmn{M2M}} \propto \Upsilon _{\rmn{Sch}} ^{1.18 \pm 0.05}.
\end{equation}
Removing NGC 524 as an `outlier' in the M2M results, we calculate the root mean square deviation of the two sets of mass-to-light ratios as $0.31$.  Assuming the errors in both the Schwarzschild and M2M methods to be similar, we arrive at an intrinsic error in the methods of $5.9\%$ (calculated as the rms deviation divided by $\sqrt 2$ divided by the mean M2M mass-to-light ratio).  This figure agrees well with the value ($6\%$) quoted by \citet{Cappellari2006} for their comparison of Jeans and Schwarzschild models, and with the theoretical model values ($\approx 5\%$) achieved in \citet{Long2010}.

Despite differences in the two modelling methods, in general the methods are delivering, as one would hope, similar mass-to-light ratios for a galaxy.  However, it is apparent from Figure \ref{fig:mlplot15} that, for the sample of galaxies analysed, either the M2M method is slightly over-estimating the mass-to-light ratios, or conversely, that Schwarzschild's method is under-estimating them.

\begin{figure*}
		\centering
		\includegraphics[width=75mm]{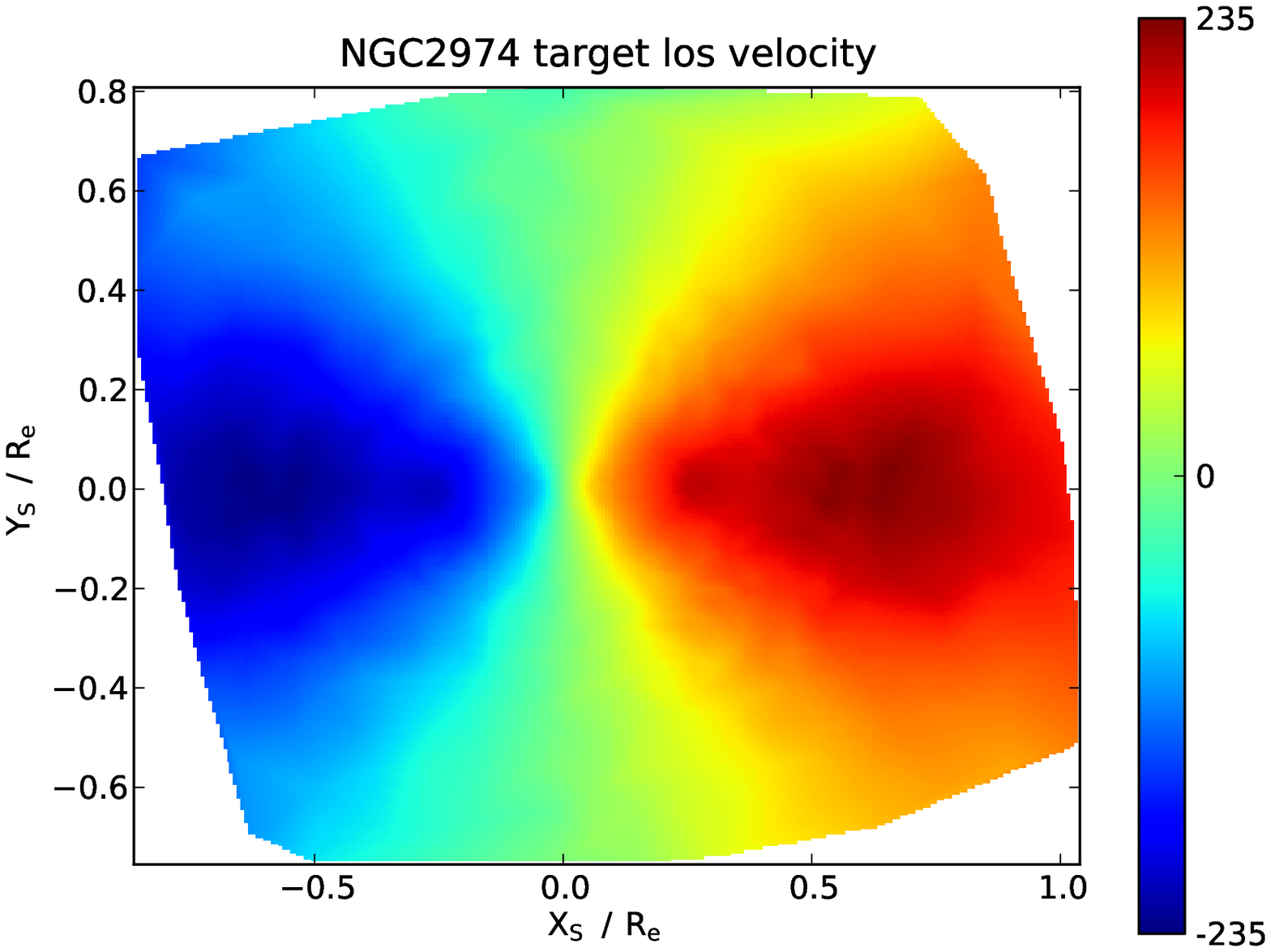}
		\includegraphics[width=75mm]{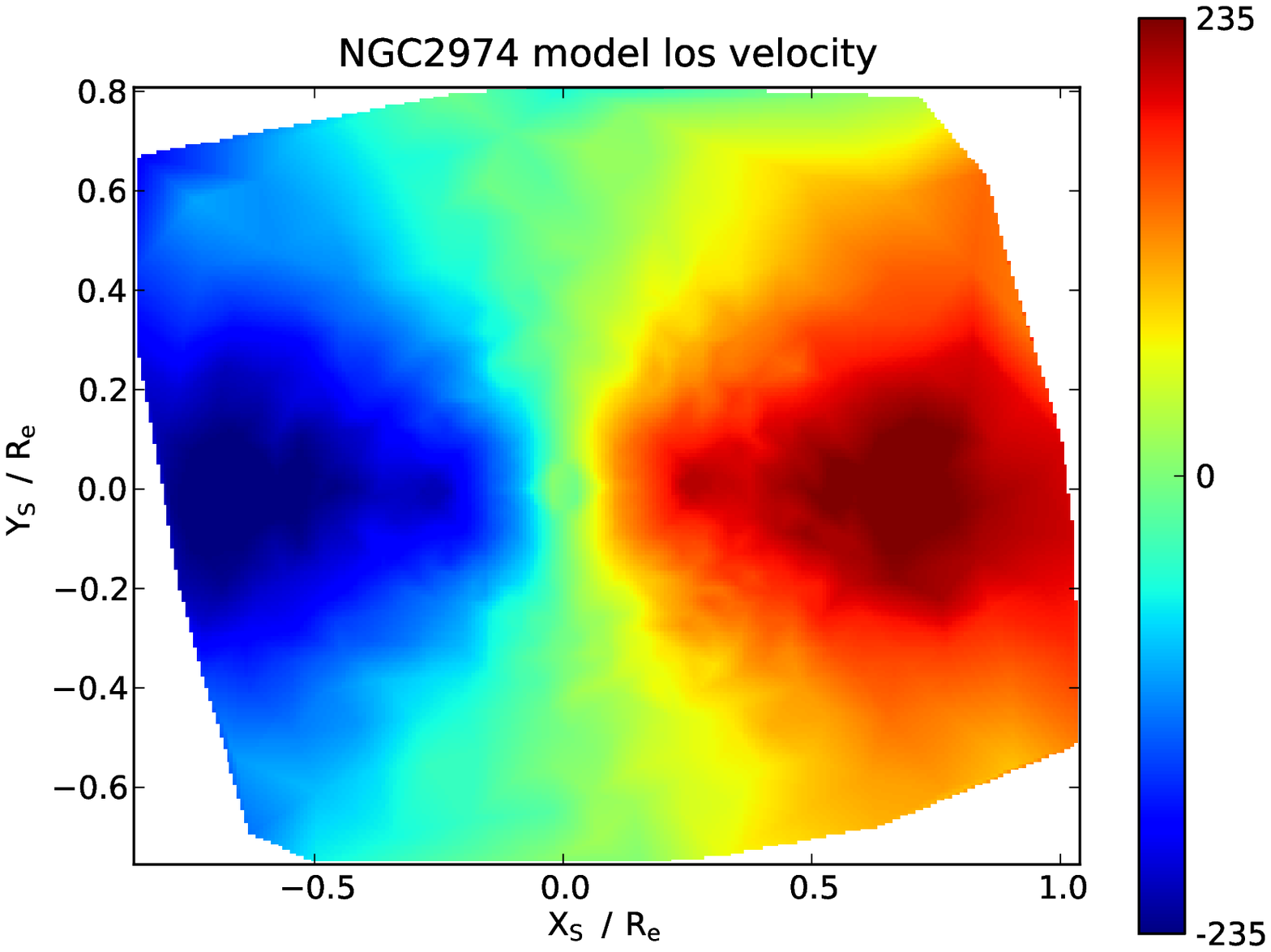}\\
		\includegraphics[width=75mm]{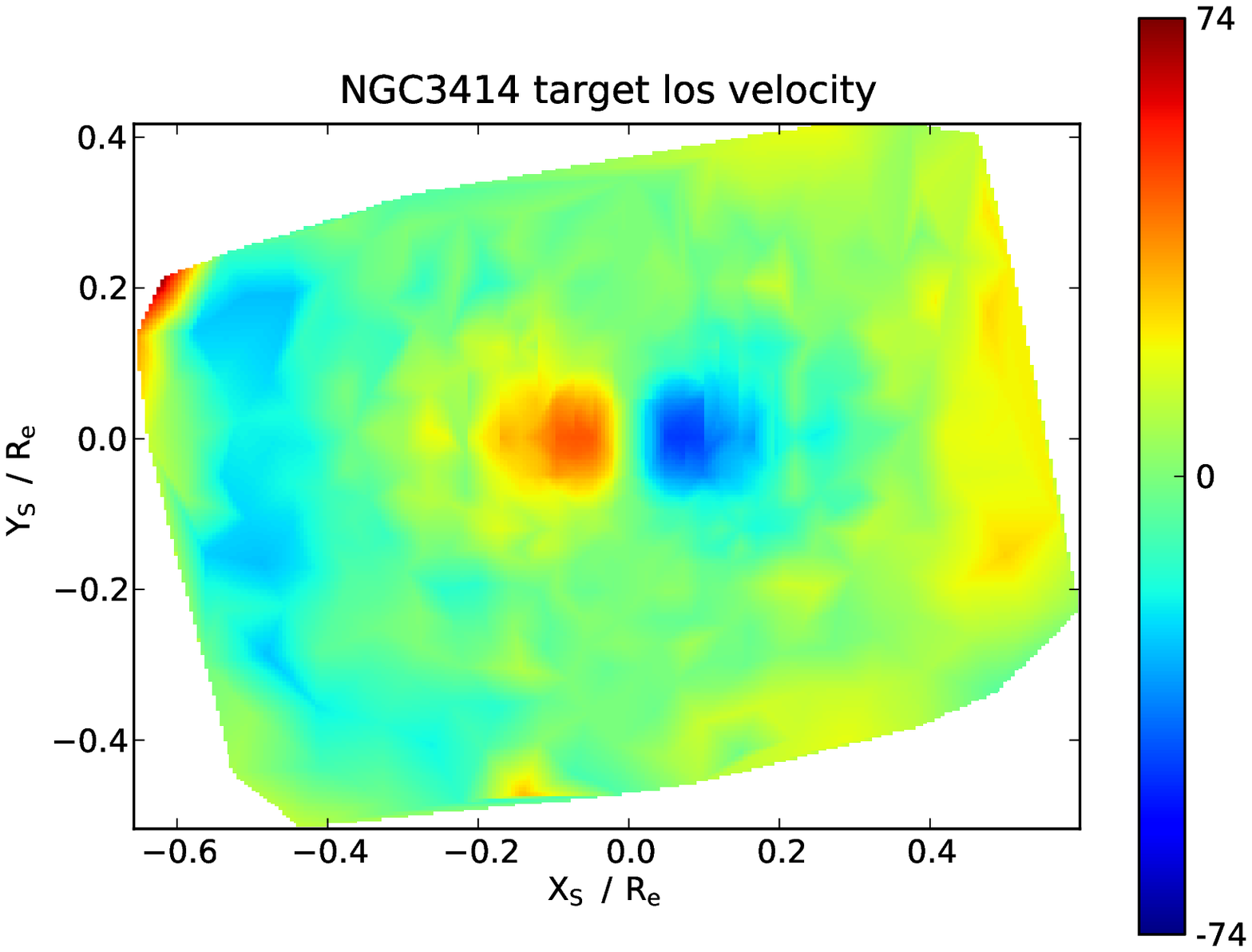}
		\includegraphics[width=75mm]{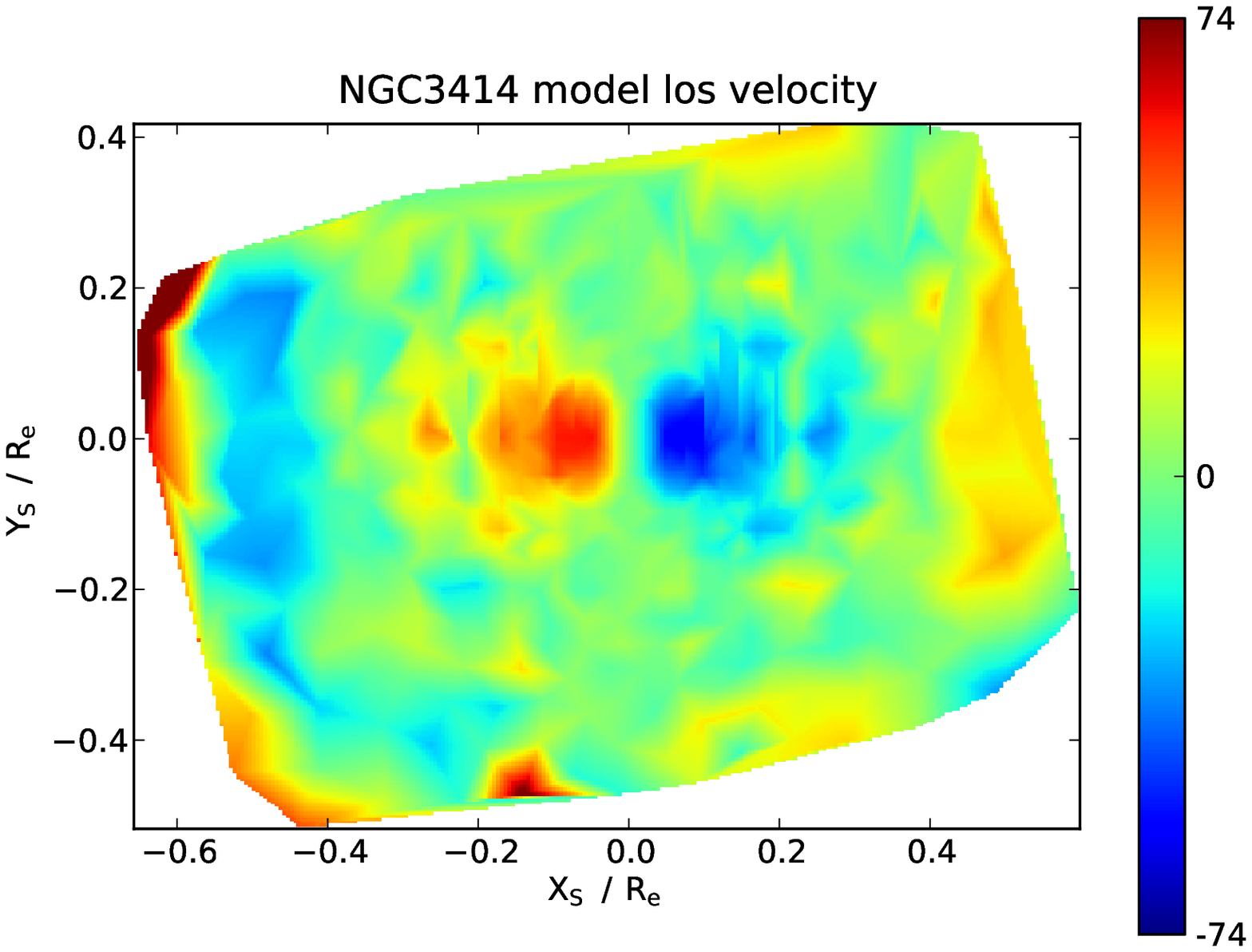}\\
		\includegraphics[width=75mm]{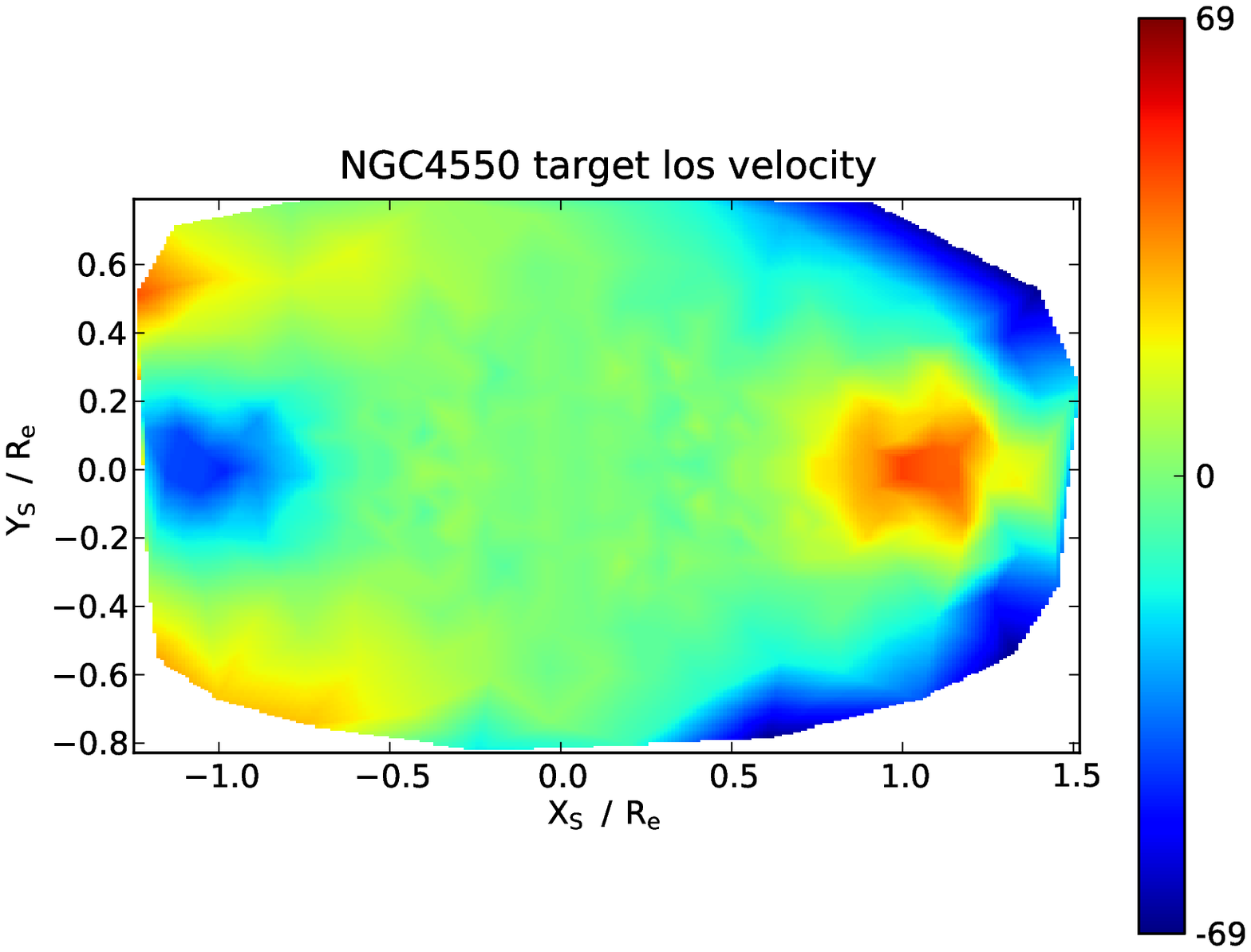}
		\includegraphics[width=75mm]{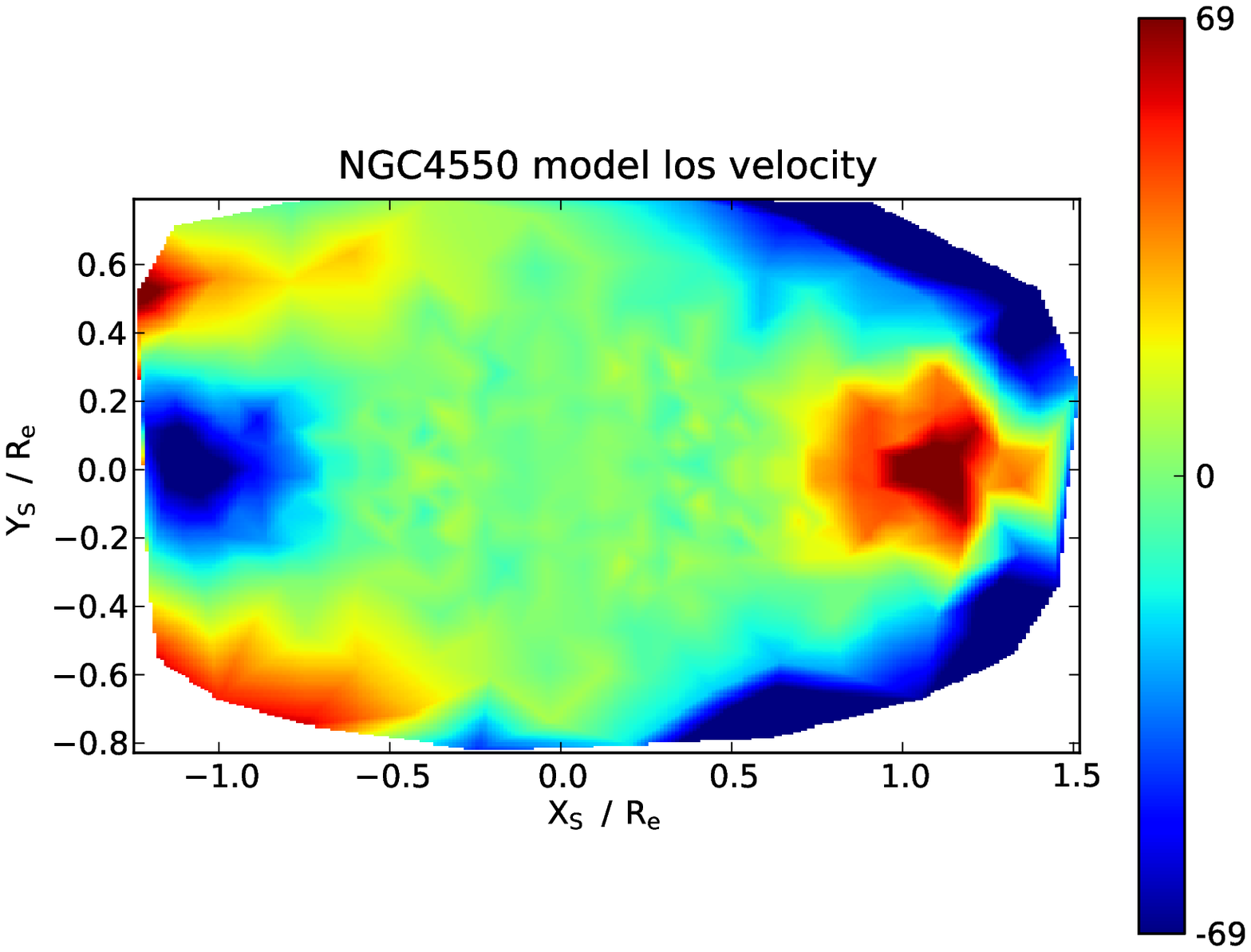}\\
		\includegraphics[width=75mm]{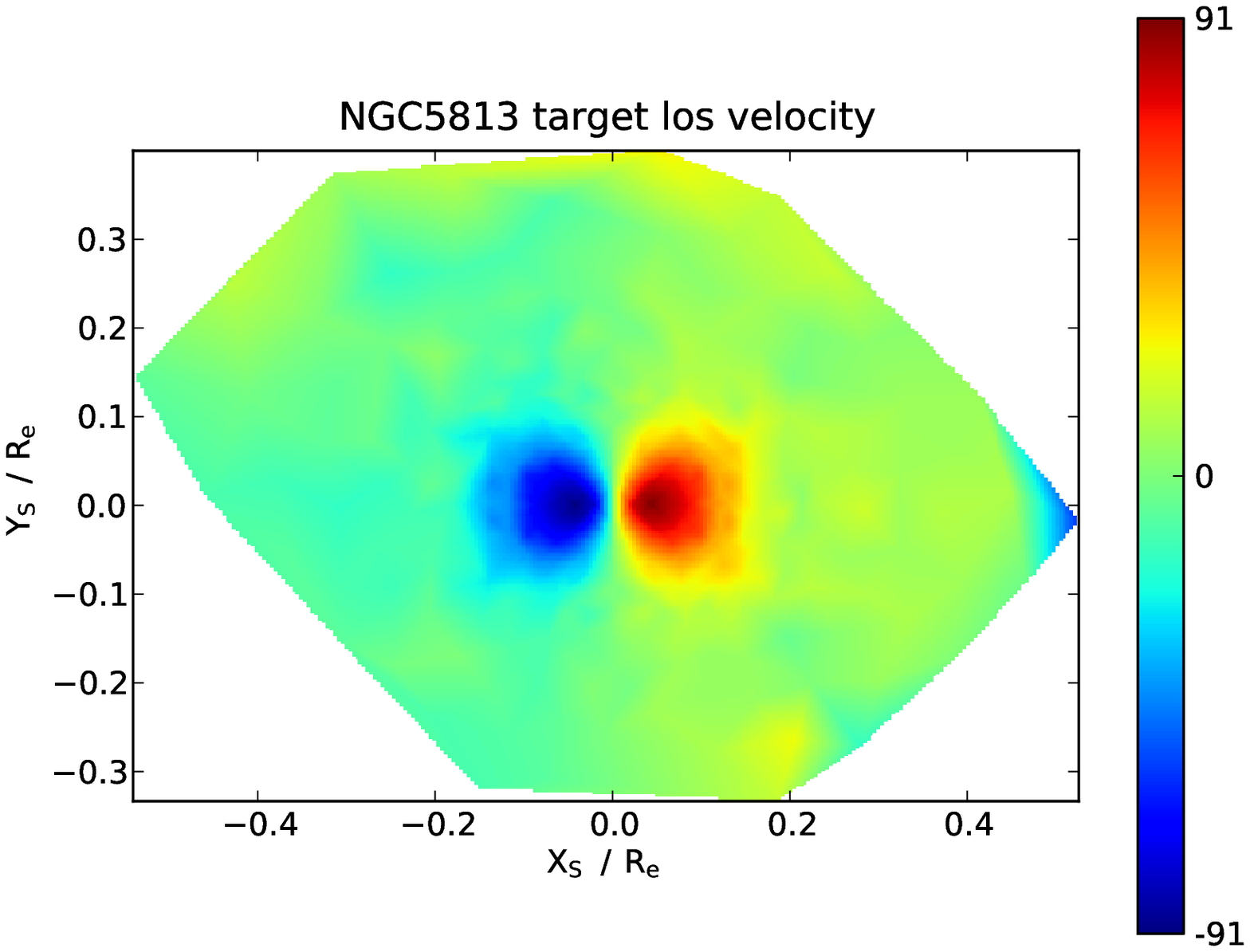}
		\includegraphics[width=75mm]{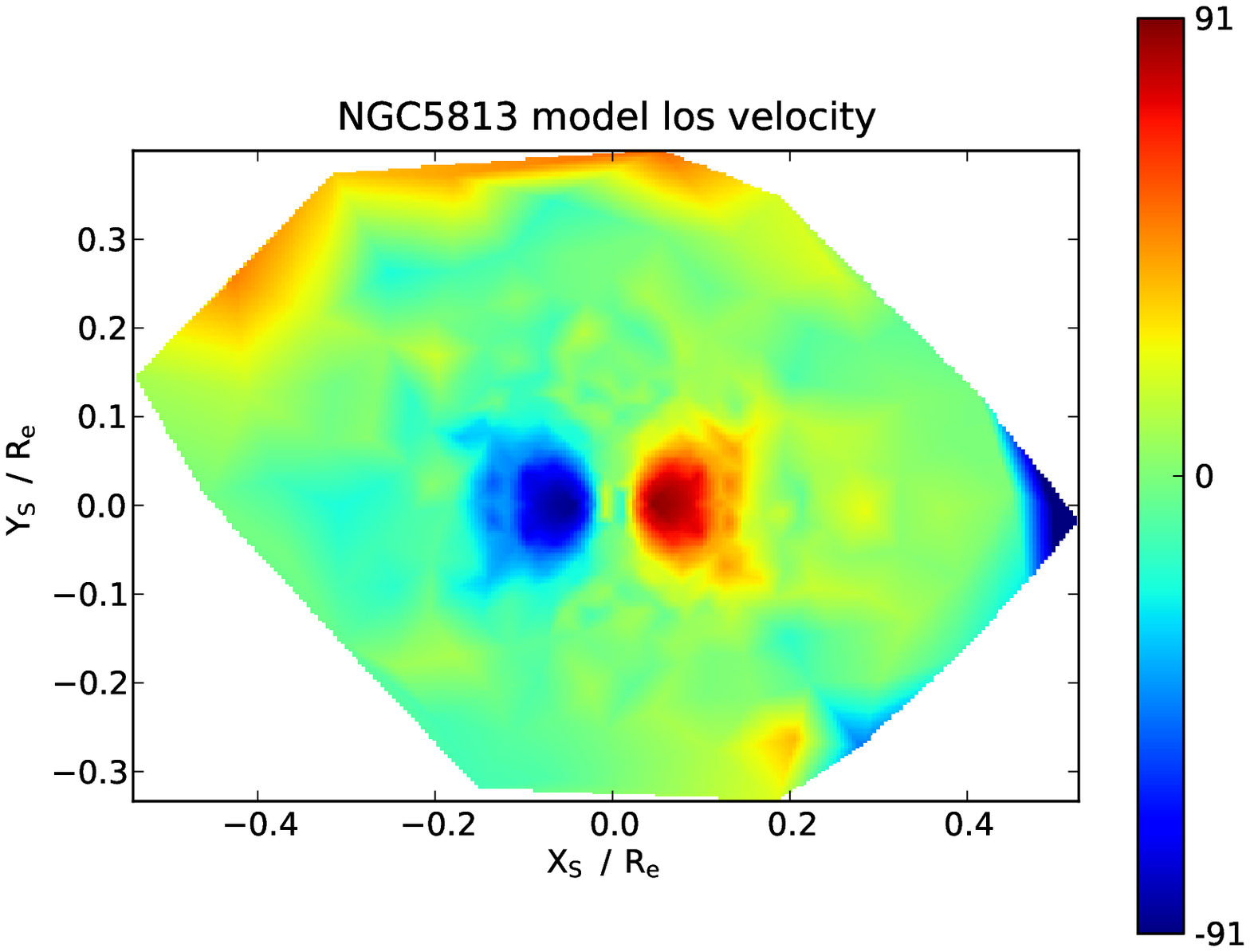}\\
		\caption[Galaxy velocity maps]{Galaxy velocity maps for NGC 2974 (inclined, elliptical, fast rotator), NGC 3414 (edge-on, lenticular, slow rotator with a counter rotating core), NGC 4550 (inclined, lenticular, fast rotator with counter rotating disc) and NGC 5813 (edge-on, elliptical, slow rotator with a kinematically distinct core) showing the target mean line-of-sight velocity and the model produced versions. See Table \ref{tab:galaxysample} for more data on the individual galaxies.  The velocity units are $\rmn{km\ s^{-1}}$.}
\label{fig:galaxyvelocity}
\end{figure*}

\begin{figure*}
		\centering
		\includegraphics[width=75mm]{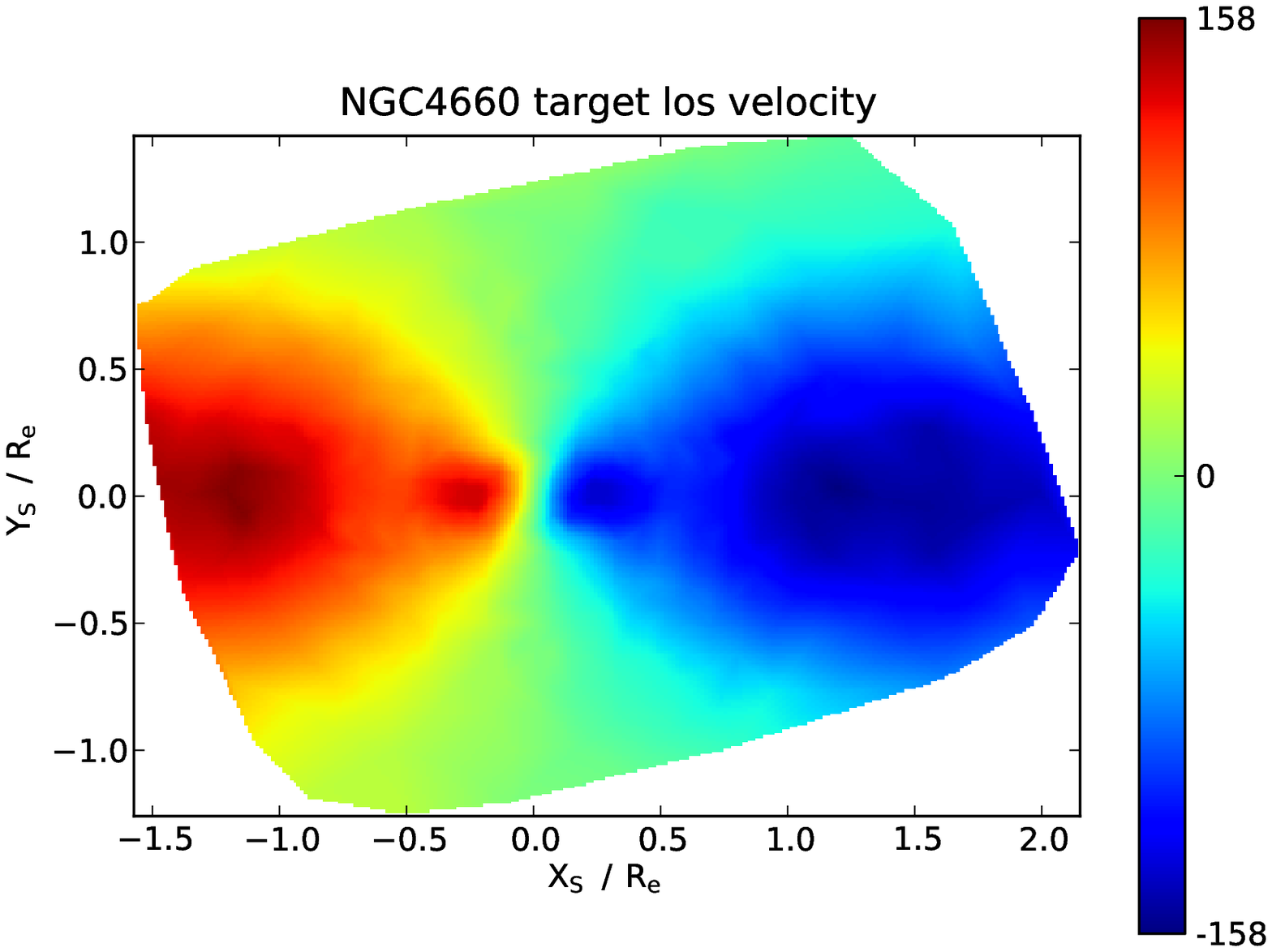}
		\includegraphics[width=75mm]{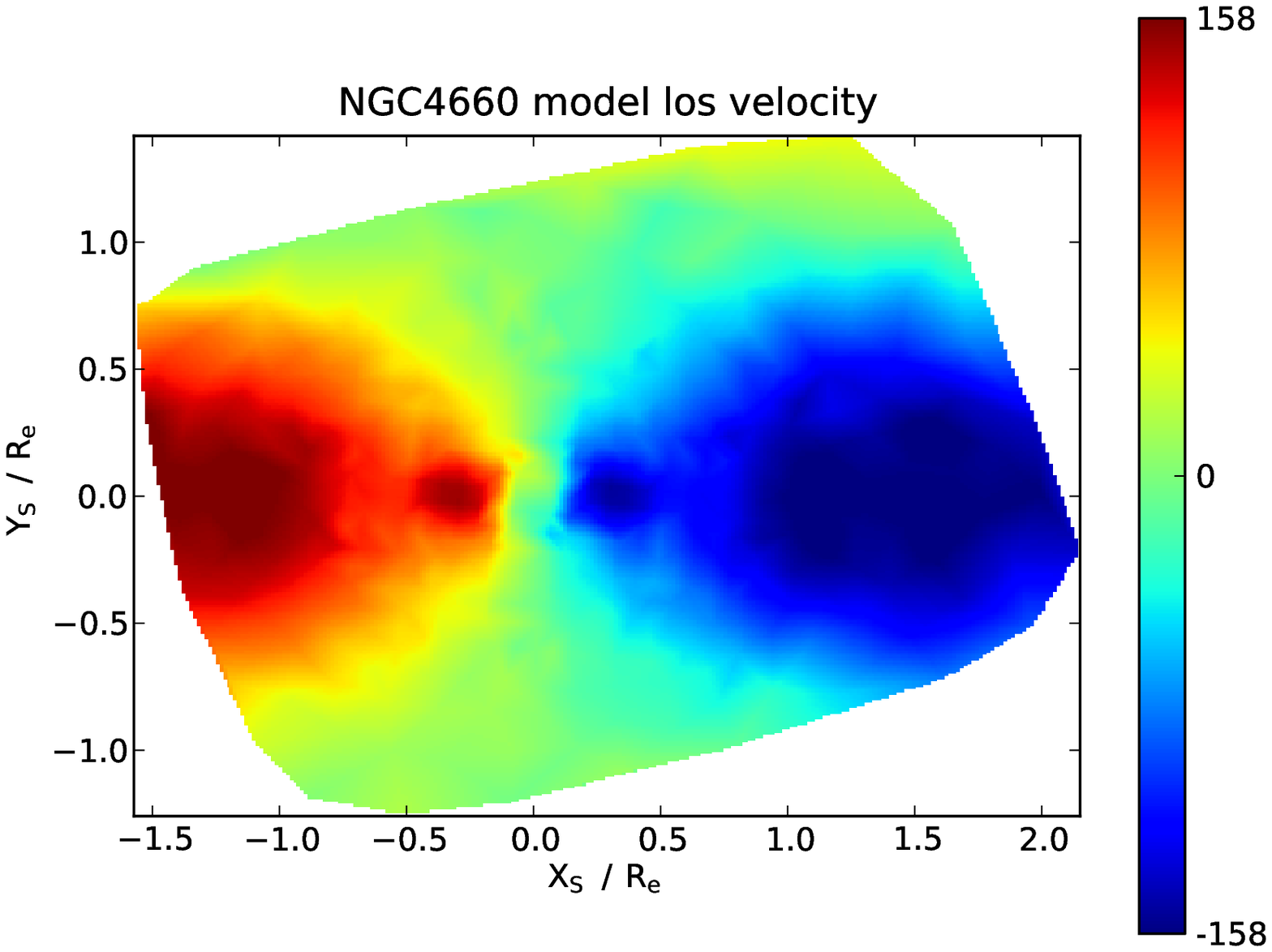}\\
		\includegraphics[width=75mm]{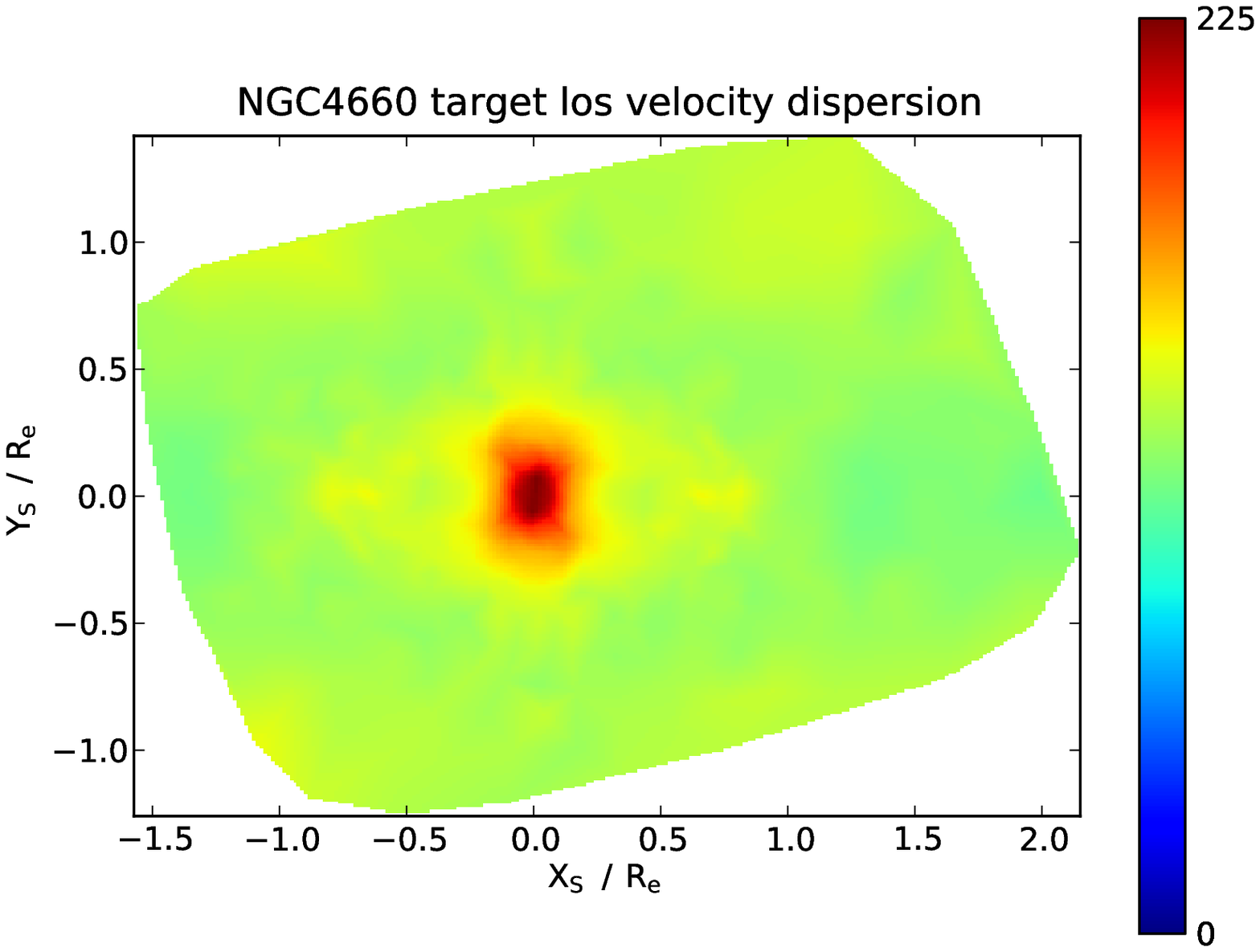}
		\includegraphics[width=75mm]{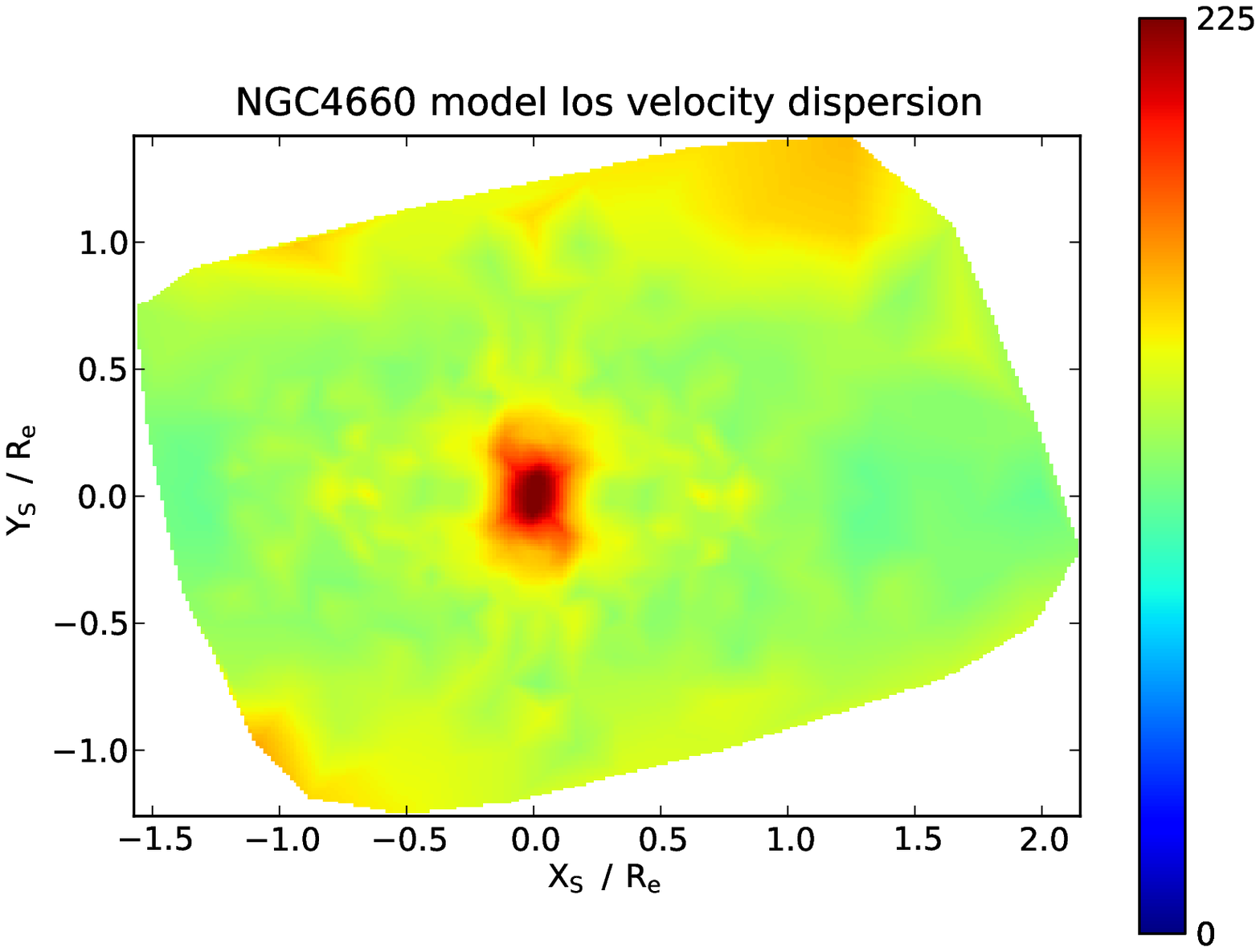}\\
		\includegraphics[width=75mm]{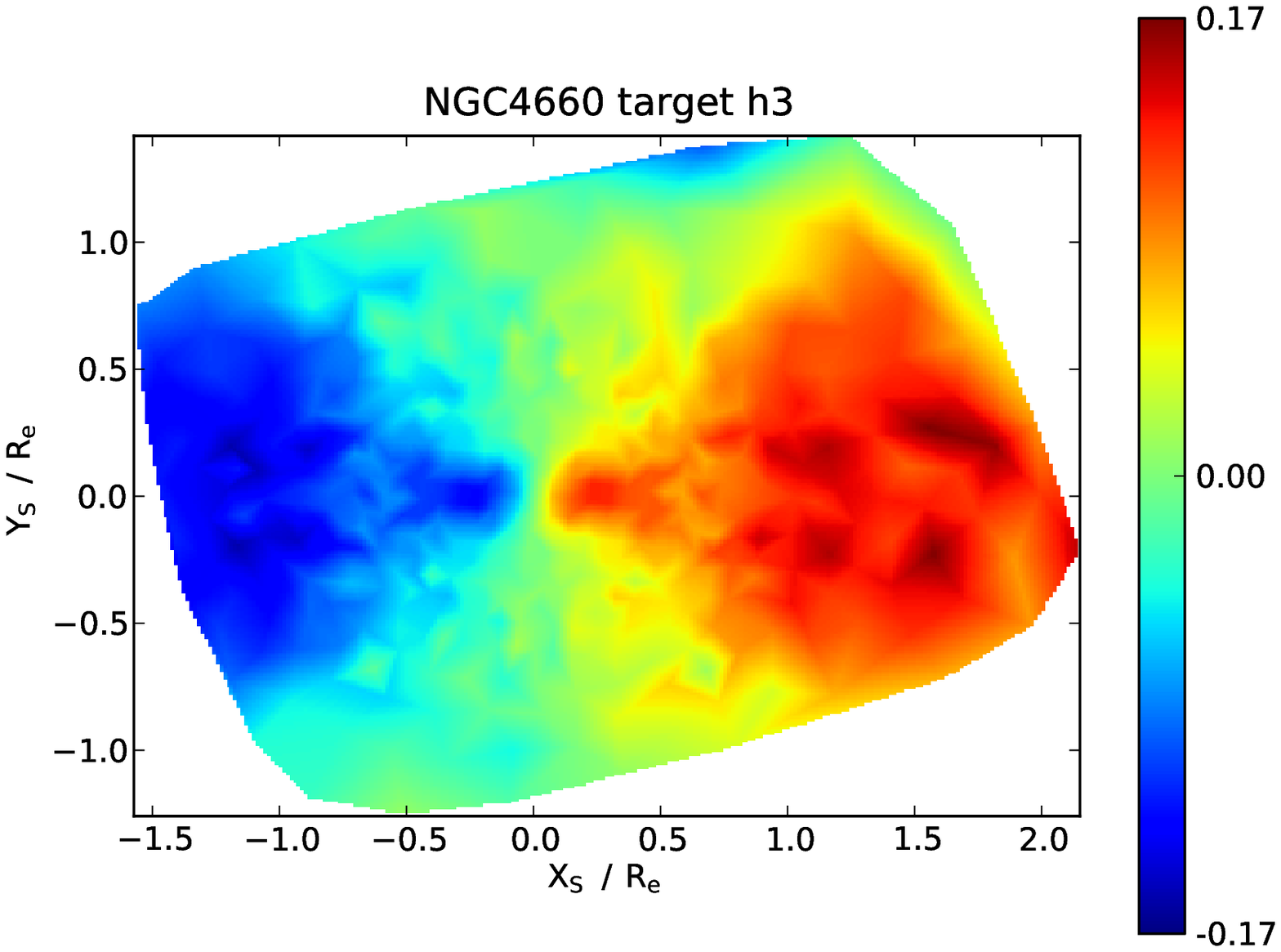}
		\includegraphics[width=75mm]{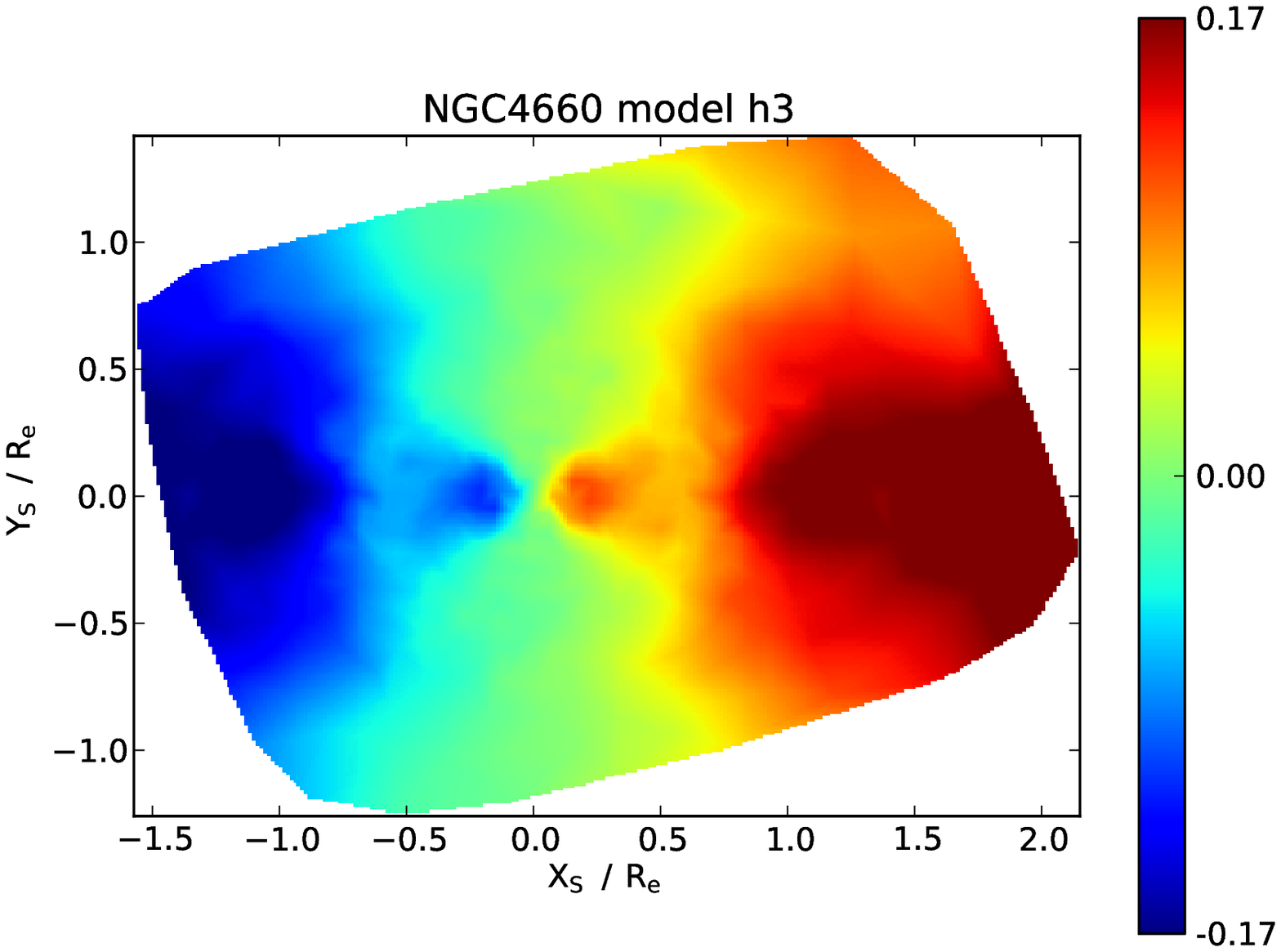}\\
		\includegraphics[width=75mm]{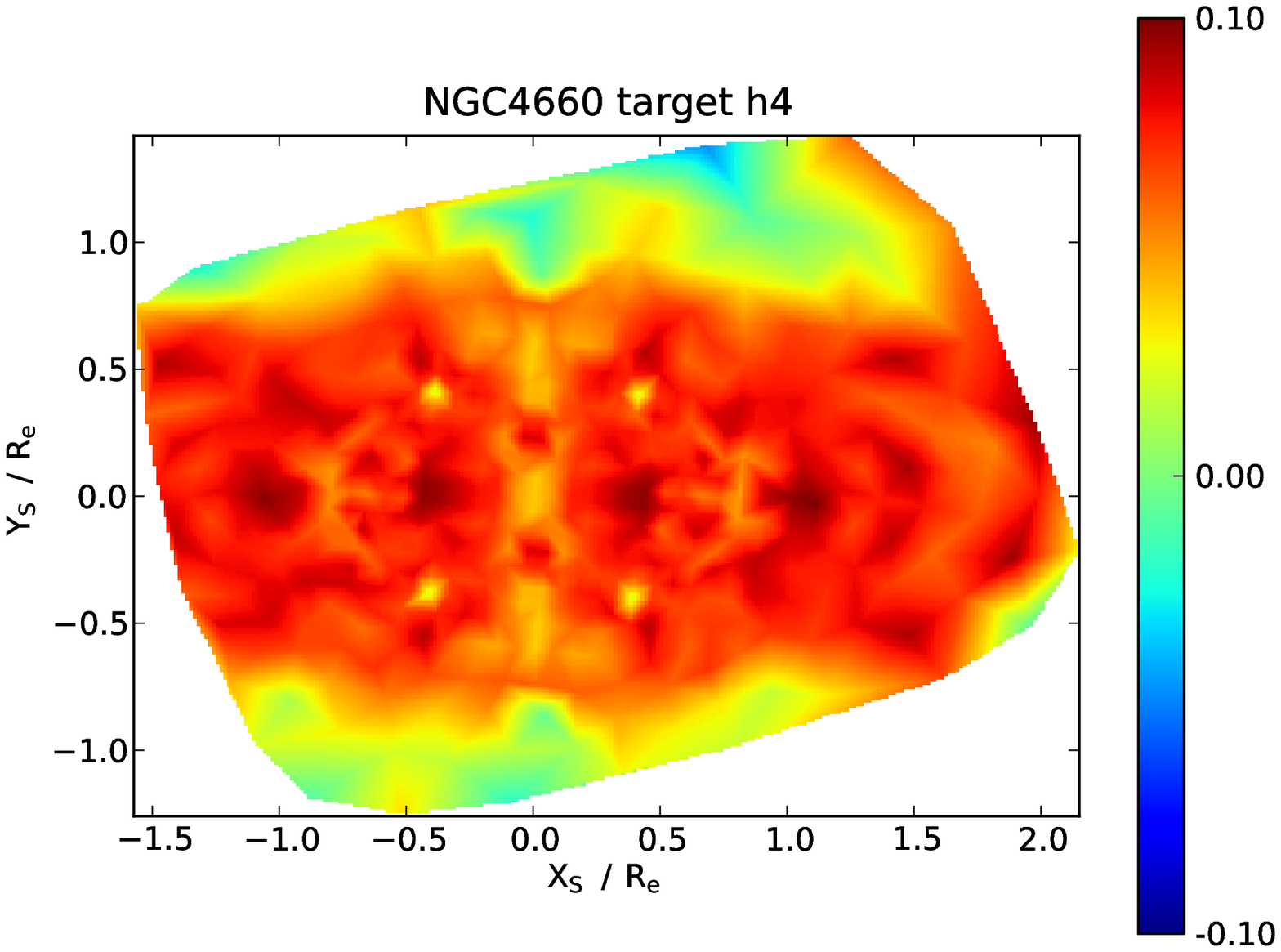}
		\includegraphics[width=75mm]{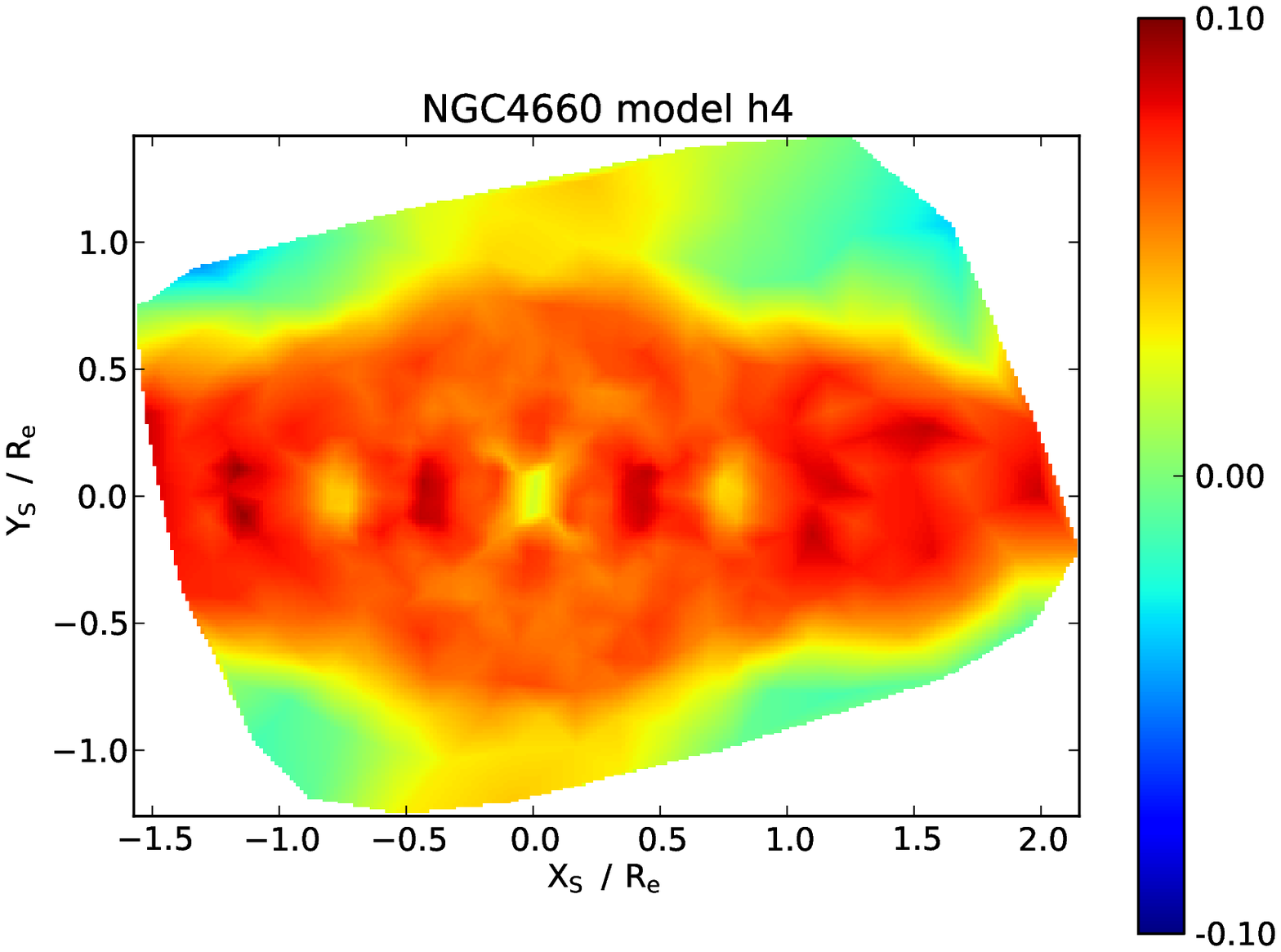}\\
		\caption[NGC 4660 observable maps]{The complete set of target and model observable maps for NGC 4660.}
\label{fig:galaxy4660}
\end{figure*}

Having determined an estimate for the galaxies' mass-to-light ratios, we perform a further set of modelling runs at those ratios using $10^6$ particles in order to investigate how well the observables are reproduced and to calculate the global anisotropy parameters (see section \ref{sec:anisotropy}). At the estimated mass-to-light ratios, weight convergence is good with $\approx 98\%$ of particles having converged weights.  The $\chi ^2$ per degree of freedom values for the constraining observables (surface brightness and the Gauss-Hermite coefficients $h_1$ to $h_6$) are generally less than $1$.  The values for the calculated observables (mean line-of-sight velocity and velocity dispersion) are similarly so.  More detail is given in Table \ref{tab:chi2anal}.  In Figure \ref{fig:galaxyvelocity}, we show the observed and model line-of-sight velocity maps for a selection of $4$ galaxies (NGC 2974, NGC 3414, NGC 4550 and NGC 5813) covering the major galaxy types (elliptical and lenticular), rotation (fast and slow), orientation (inclined to the line-of-sight and edge-on) and core features.  In particular, we note that the M2M method is able to reproduce kinematically distinct cores and counter rotating cores.  For completeness, Figure \ref{fig:galaxy4660} contains a complete set of observable maps (velocity, dispersion, $h_3$, and $h_4$) for NGC 4660. Overall, reproduction is satisfactory.

\begin{table}
	\centering
	\caption{$\chi^2$ per degree of freedom analysis - all galaxies}
	\label{tab:chi2anal}
	\begin{tabular}{ccccc}
		\hline
		Observable & Minimum & Maximum & Mean & Median \\
		  & $\chi ^2 / \rmn{dof}$ & $\chi ^2 / \rmn{dof}$ & $\chi ^2 / \rmn{dof}$ & $\chi ^2 / \rmn{dof}$\\
		\hline
		sb        & $0.05$ & $0.64$ & $0.30$ & $0.25$\\
		$h_1$     & $0.05$ & $0.96$ & $0.35$ & $0.27$\\
		$h_2$     & $0.39$ & $1.22$ & $0.73$ & $0.75$\\
		$h_3$     & $0.19$ & $2.17$ & $0.58$ & $0.50$\\
		$h_4$     & $0.36$ & $1.20$ & $0.66$ & $0.61$\\
		$h_5$     & $0.000$ & $0.023$ & $0.004$ & $0.001$\\
		$h_6$     & $0.002$ & $0.021$ & $0.009$ & $0.008$\\
		$v _m$ & $0.05$ & $0.97$ & $0.35$ & $0.27$\\
		$\sigma _m$  & $0.33$ & $2.31$ & $0.78$ & $0.70$\\
		\hline
	\end{tabular}
	
\medskip
The M2M constraining observables are surface brightness (sb) and $h1$ to $h6$. The model mean line-of-sight velocity ($v_m$) and velocity dispersion ($\sigma _m$) are calculated as in equations \ref{eqn:modelv} and \ref{eqn:modelsigma}. The maximum values of $h_3$ and $v _m$ come from the modelling of NGC 5845 and the maximum of $\sigma _m$ from NGC 3156.
\end{table}

\section{Anisotropy Parameters}\label{sec:anisotropy}

\subsection{Theory}
The global anisotropy parameter $\delta$ is given in \citet{BT2008} \textsection 4.8 and \citet{SauronX2007} as
\begin{equation}
	\delta = 1 - \frac{\Pi _{zz}}{\Pi _{xx}}
\end{equation}
where $\Pi _{ij}$ is from the kinetic energy due to random stellar motion,  $z$ indicates the symmetry axis of an axisymmetric galaxy and $x$ is some fixed direction orthogonal to it.
$\Pi _{kk}$ is defined as
\begin{equation}
	\Pi _{kk} = \int \rho \sigma _k ^2 \; d^3 \bmath{x}
\end{equation}
where $\rho$ is the mass density and $\sigma _k$ the velocity dispersion in direction $k$. 
The equivalent for M2M modelling purposes, calculated using the particle weights and binning particle data into $J$ bins, is
\begin{equation}
	\Pi _{kk} = \Upsilon L \sum _j ^J W_j \sigma _{k,j} ^2
\end{equation}
where $\Upsilon$ is the mass-to-light ratio of the galaxy, $L$ is the model luminosity, $\sigma _{k,j}$ is the mean luminosity weighted velocity dispersion in direction $k$ in bin $j$, and $W_j$, the sum of the particles weights in bin $j$, is given by
\begin{equation}
	W _j = \sum _i ^N w_i \; \delta (i \in j).
\end{equation}

\citet{SauronX2007} introduce two further parameters, $\beta$ and $\gamma$.  Using cylindrical polar coordinates $(R, \phi, z)$,
\begin{equation}
	\beta = 1 - \frac{\Pi _{zz}}{\Pi _{RR}}
\end{equation}
and
\begin{equation}
	\gamma = 1 - \frac{\Pi _{\phi \phi}}{\Pi _{RR}}.
\end{equation}
The global anisotropy parameter $\delta$ is then calculated as
\begin{equation}
	\delta = \frac{2 \beta - \gamma}{2 - \gamma}.
\end{equation}
As noted in \citet{SauronX2007}, $\beta$ describes the global shape of the velocity dispersion tensor in the $(v_R, v_z)$ plane, and $\gamma$ the shape in a plane orthogonal to $v_z$.

The three anisotropy parameters described so far, whilst still applicable to spherical galaxies, are more appropriate to axisymmetric galaxies.  \citet{SauronX2007} describe a further parameter, $\beta_r$, to measure the anisotropy of (near) spherical galaxies.
\begin{equation}
	\beta _r = 1 - \frac{\Pi _{\theta \theta} + \Pi _{\phi \phi}}{2\;\Pi _{rr}}
\end{equation}
where $(r, \theta, \phi)$ are spherical coordinates.

\begin{figure*}
		\centering
		\includegraphics[width=84mm]{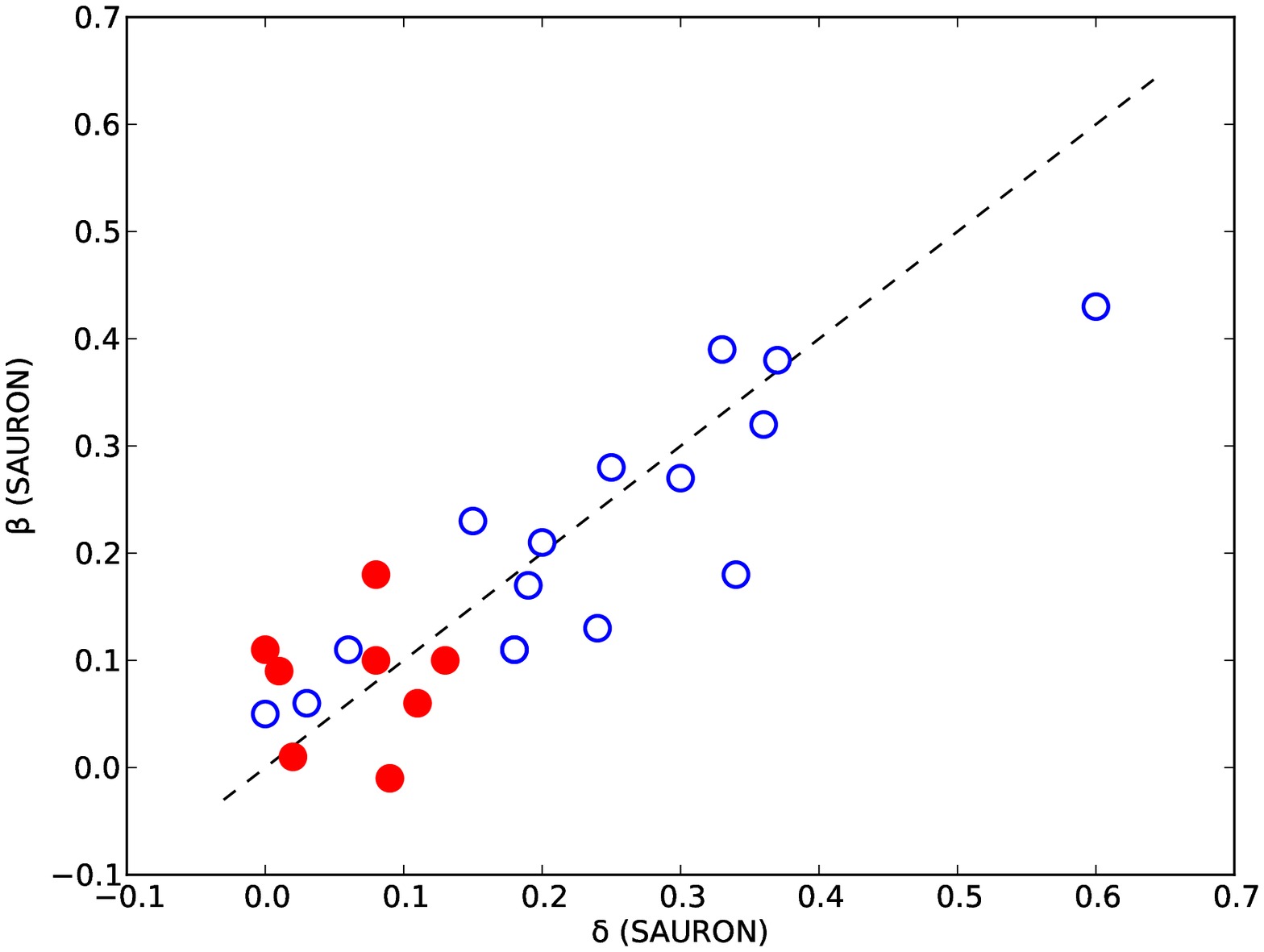}
		\includegraphics[width=84mm]{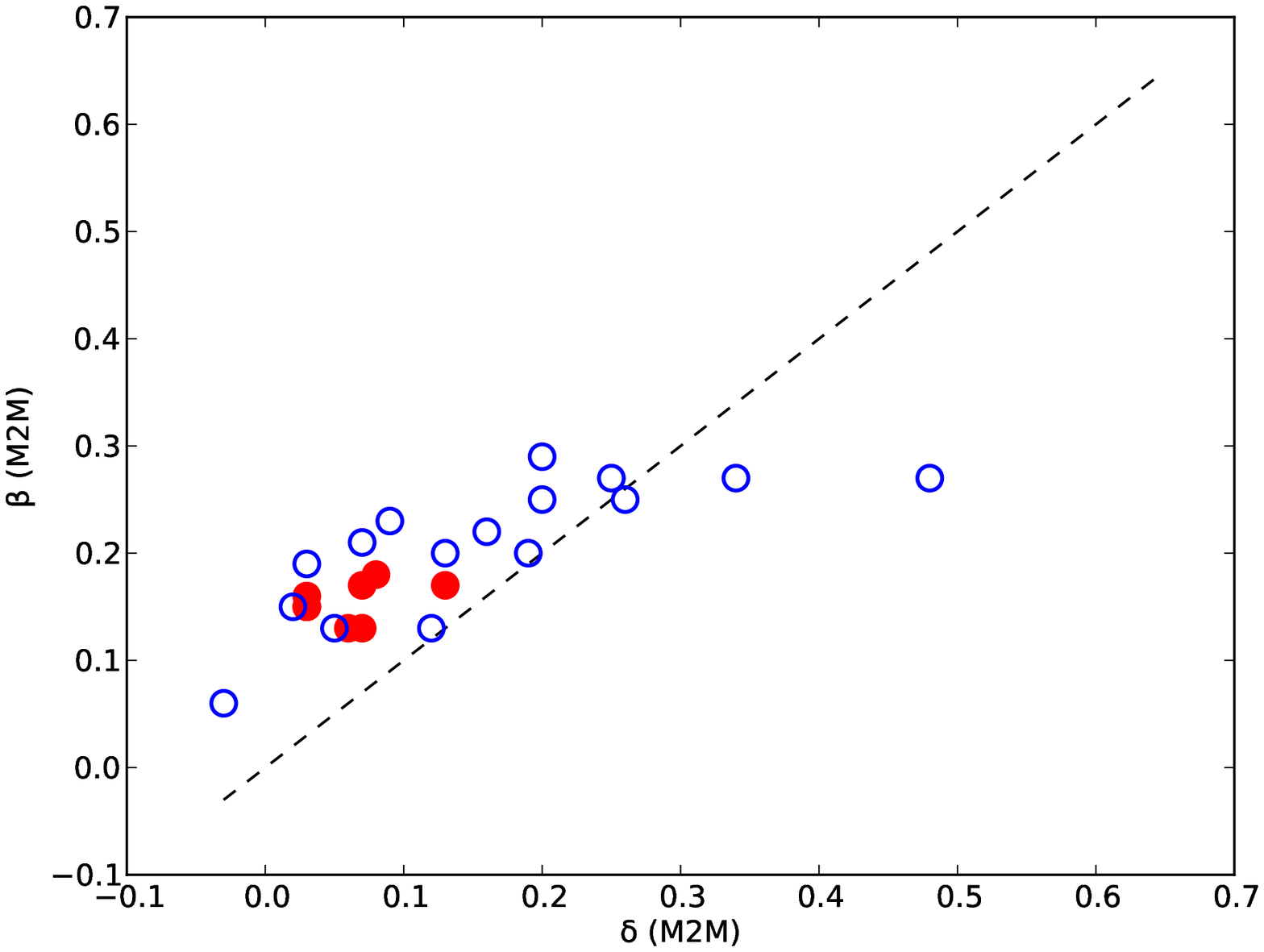}\\
		\includegraphics[width=84mm]{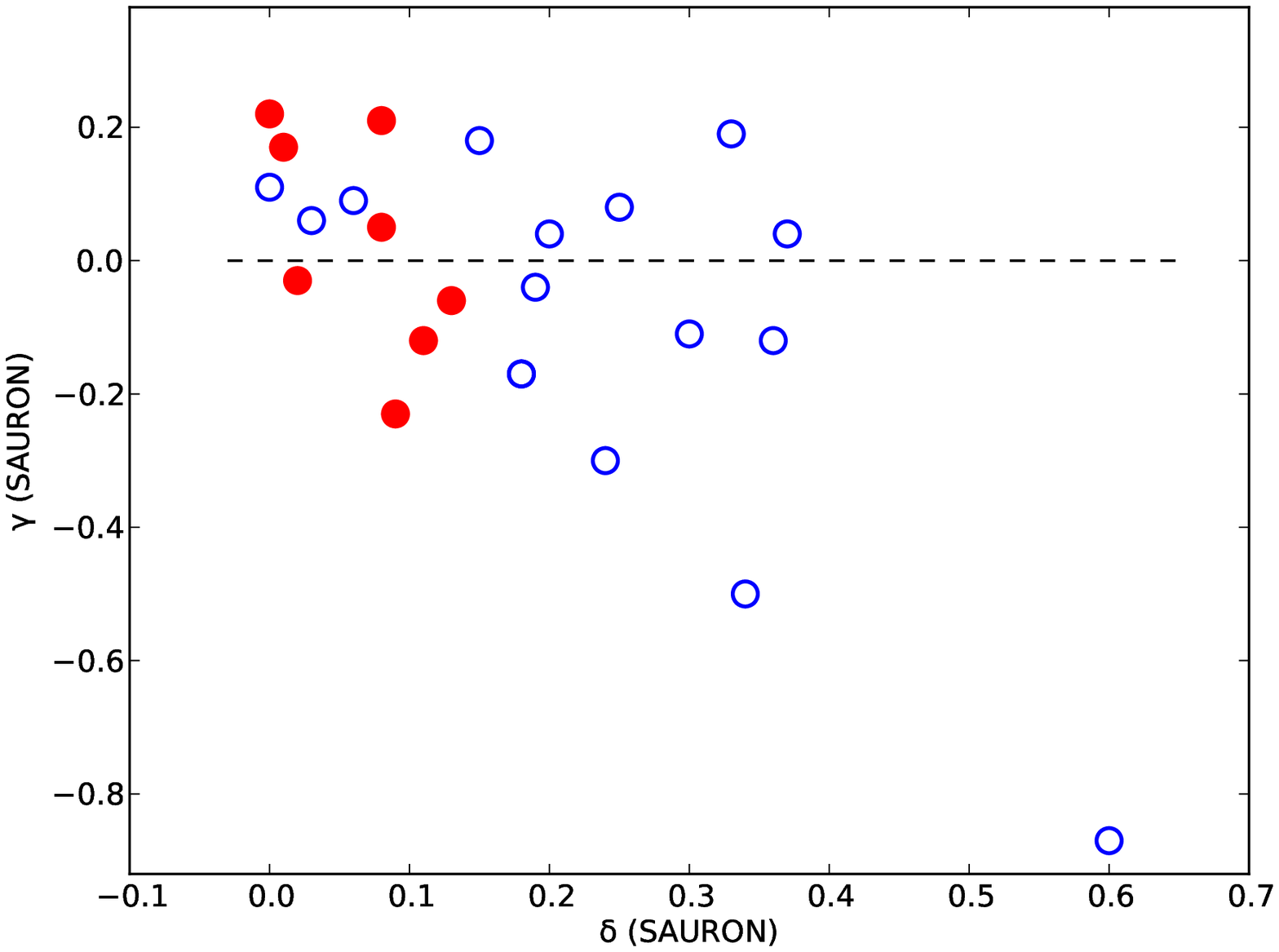}
		\includegraphics[width=84mm]{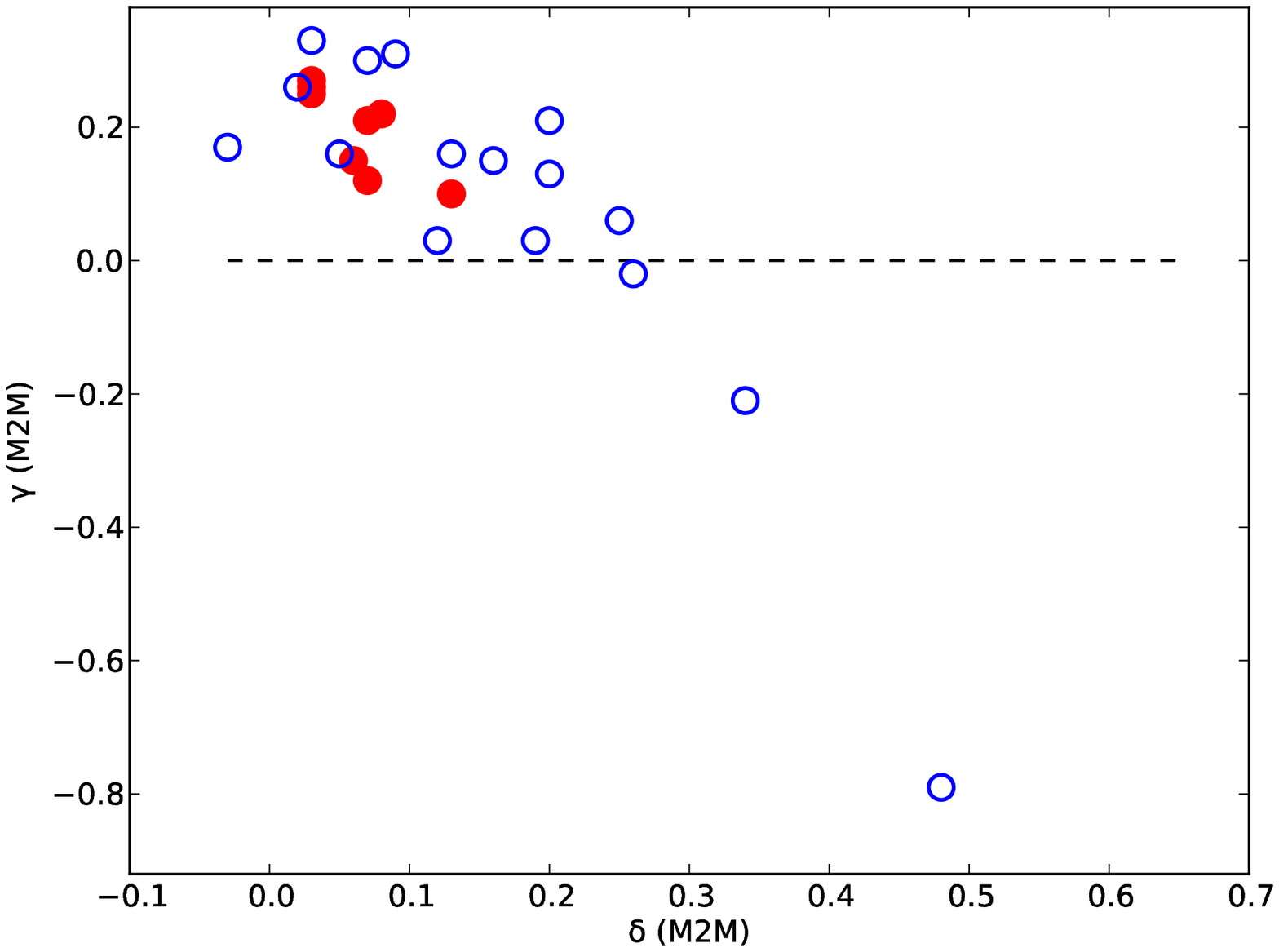}\\
		\includegraphics[width=84mm]{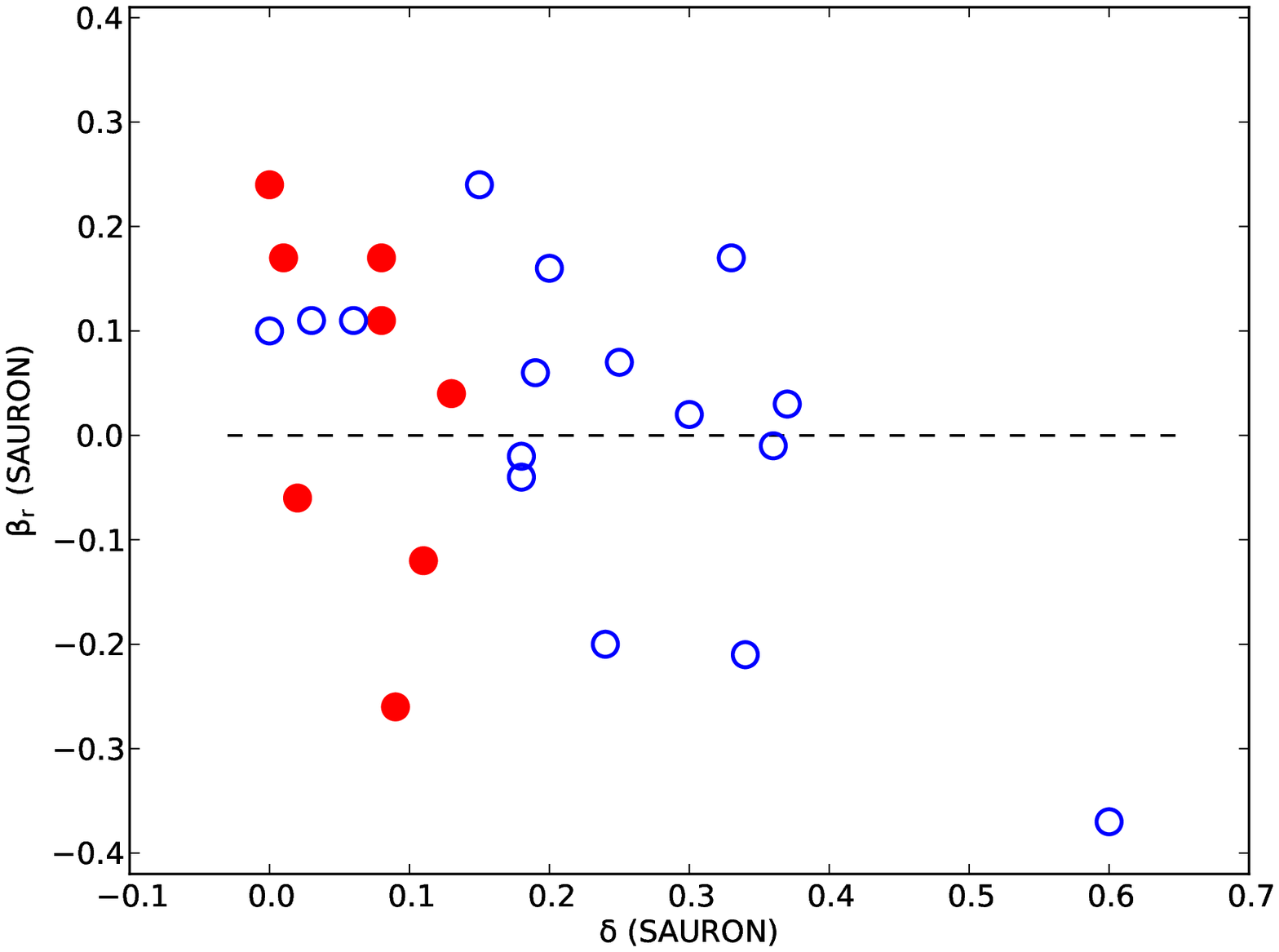}
		\includegraphics[width=84mm]{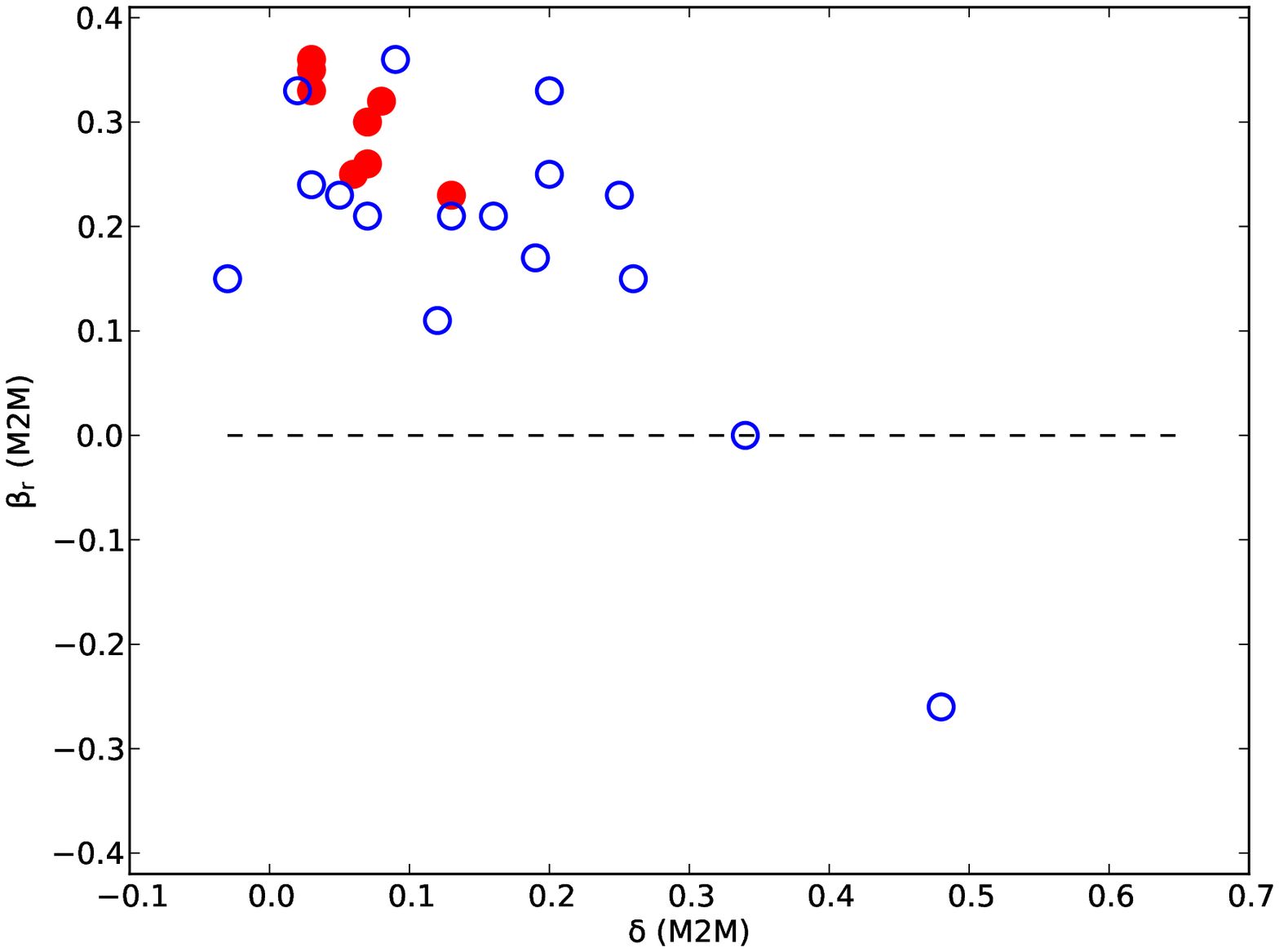}
		\caption[Anisotropy parameters]{Anisotropy parameters $\beta$, $\gamma$ and $\beta _r$ for the individual galaxies  plotted against $\delta$. Slow rotating galaxies are plotted with red solid markers.  The M2M plots are in the right hand column. Plots from \citet{SauronX2007} (left hand column) have been included for comparison purposes.  The black dashed lines separate the plot areas into radial (upper or upper left) and tangential (lower or lower right) anisotropy regimes.  It is clear that fewer M2M models exhibit tangential velocity dispersion anisotropy and that the slow rotating galaxies are more tightly clustered in parameter space than their SAURON equivalents.}
		\label{fig:aniparams}
\end{figure*}

For all parameters, the radial and tangential velocity dispersion anisotropy regimes are shown in Figure 6.  For the directions indicated within the definitions of the parameters, a zero parameter value indicates isotropy and a positive value, a radial bias to the velocity dispersion.

\subsection{M2M modelling}
For each galaxy, we perform a M2M modelling run using the mass-to-light ratio determined in section \ref{sec:ml}. We calculate the anisotropy parameters by binning the end of run particle velocity data on an $(R, z)$ grid for $\beta$ and $\gamma$, and for $\beta _r$, we bin the data radially. Given we are using the particle data directly, the number of particles is increased to $1 \times 10^6$. Similarly to \citet{SauronX2007}, we only include particles in the calculation which are currently within $25$ arcsec spherical radius of the galactic centre.

\begin{table*}
	\centering
	\caption{Comparison between SAURON and M2M anisotropy parameters}
	\label{tab:anicomp}
	\begin{tabular}{lccrrrrrrrr}
		\hline
 & & & \multicolumn{4}{c}{\citet{SauronX2007}} & \multicolumn{4}{c}{From M2M models} \\
Galaxy & Inclination & Fast & $\beta _r$ & $\beta$ & $\gamma$ & $\delta$ & $\beta _r$ & $\beta$ & $\gamma$ & $\delta$ \\
       & (deg) & Rotator & & & & & & & & \\
		\hline
NGC 524   & 19 & yes  & 0.06  & 0.17 & -0.04 & 0.19   & 0.00  & 0.27 & -0.21  & 0.34 \\
NGC 821   & 90 & yes  & 0.16  & 0.21 &  0.04 & 0.20   & 0.25  & 0.25 &  0.13  & 0.20 \\
NGC 2974  & 57 & yes & -0.20  & 0.13 & -0.30 & 0.24   & 0.11  & 0.13  & 0.03  & 0.12 \\
NGC 3156  & 68 & yes  & 0.17  & 0.39  & 0.19 & 0.33   & 0.24  & 0.19  & 0.33  & 0.03 \\
NGC 3377  & 90 & yes  & 0.07  & 0.28  & 0.08 & 0.25   & 0.21  & 0.22  & 0.15  & 0.16 \\
NGC 3379  & 90 & yes &  0.11  & 0.06  & 0.06 & 0.03   & 0.33  & 0.15  & 0.26  & 0.02 \\
NGC 3414  & 90 & no  & -0.12  & 0.06 & -0.12 & 0.11   & 0.25  & 0.13  & 0.15  & 0.06 \\
NGC 3608  & 90 & no  &  0.04  & 0.10 & -0.06 & 0.13   & 0.23  & 0.17  & 0.10  & 0.13 \\
NGC 4150  & 52 & yes & -0.01  & 0.32 & -0.12 & 0.36   & 0.15  & 0.06  & 0.17  & -0.03 \\
NGC 4278  & 90 & yes & -0.02  & 0.11 & -0.17 & 0.18   & 0.33  & 0.29  & 0.21  & 0.20 \\
NGC 4374  & 90 & no   & 0.11  & 0.10 &  0.05 & 0.08   & 0.30  & 0.17  & 0.21  & 0.07 \\
NGC 4458  & 90 & no  & -0.26 & -0.01 & -0.23 & 0.09   & 0.26  & 0.13  & 0.12  & 0.07 \\
NGC 4459  & 47 & yes &  0.10  & 0.05  & 0.11 & 0.00   & 0.23  & 0.13  & 0.16  & 0.05 \\
NGC 4473  & 73 & yes & -0.21  & 0.18 & -0.50 & 0.34   & 0.15  & 0.25 & -0.02  & 0.26 \\
NGC 4486  & 90 & no   & 0.24  & 0.11  & 0.22 & 0.00   & 0.35  & 0.15  & 0.26  & 0.03 \\
NGC 4526  & 79 & yes &  0.11  & 0.11  & 0.09 & 0.06   & 0.21  & 0.20  & 0.16  & 0.13 \\
NGC 4550  & 84 & yes & -0.37  & 0.43 & -0.87 & 0.60  & -0.26  & 0.27 & -0.79  & 0.48 \\
NGC 4552  & 90 & no  & -0.06  & 0.01 & -0.03 & 0.02   & 0.33  & 0.16  & 0.27  & 0.03 \\
NGC 4621  & 90 & yes & -0.04  & 0.11 & -0.17 & 0.18   & 0.17  & 0.20  & 0.03  & 0.19 \\
NGC 4660  & 70 & yes  & 0.02  & 0.27 & -0.11 & 0.30   & 0.23  & 0.27  & 0.06  & 0.25 \\
NGC 5813  & 90 & no   & 0.17  & 0.18  & 0.21 & 0.08   & 0.32  & 0.18  & 0.22  & 0.08 \\
NGC 5845  & 90 & yes  & 0.24  & 0.23  & 0.18 & 0.15   & 0.36  & 0.23  & 0.31  & 0.09 \\
NGC 5846  & 90 & no   & 0.17  & 0.09  & 0.17 & 0.01   & 0.36  & 0.15  & 0.25  & 0.03 \\
NGC 7457  & 64 & yes  & 0.03  & 0.38  & 0.04 & 0.37   & 0.21  & 0.21  & 0.30  & 0.07 \\
		\hline
	\end{tabular}
	
\medskip
Comparison between the SAURON values for the anisotropy parameters $\beta _r$, $\beta$, $\gamma$ and $\delta$ taken from \citet{SauronX2007} Table 2, and the same parameters calculated from M2M models. For the $\delta$ parameter, $14$ M2M values (including all the slow rotating galaxies) are within $\pm 0.05$ of the SAURON values.
\end{table*}

Overall, we achieve reasonable (but not good) agreement with the SAURON values of the global anisotropy parameter $\delta$ with $14$ of the $24$ galaxies having M2M values equal to $\delta_{\rmn{SAURON}} \pm 0.05$.  The detailed results are captured in Table \ref{tab:anicomp}.  Figure \ref{fig:aniparams} shows $\beta$, $\gamma$ and $\beta _r$ plotted against $\delta$ and is the equivalent of \citet{SauronX2007} Figure 2.  
\begin{figure}
		\centering
		\includegraphics[width=84mm]{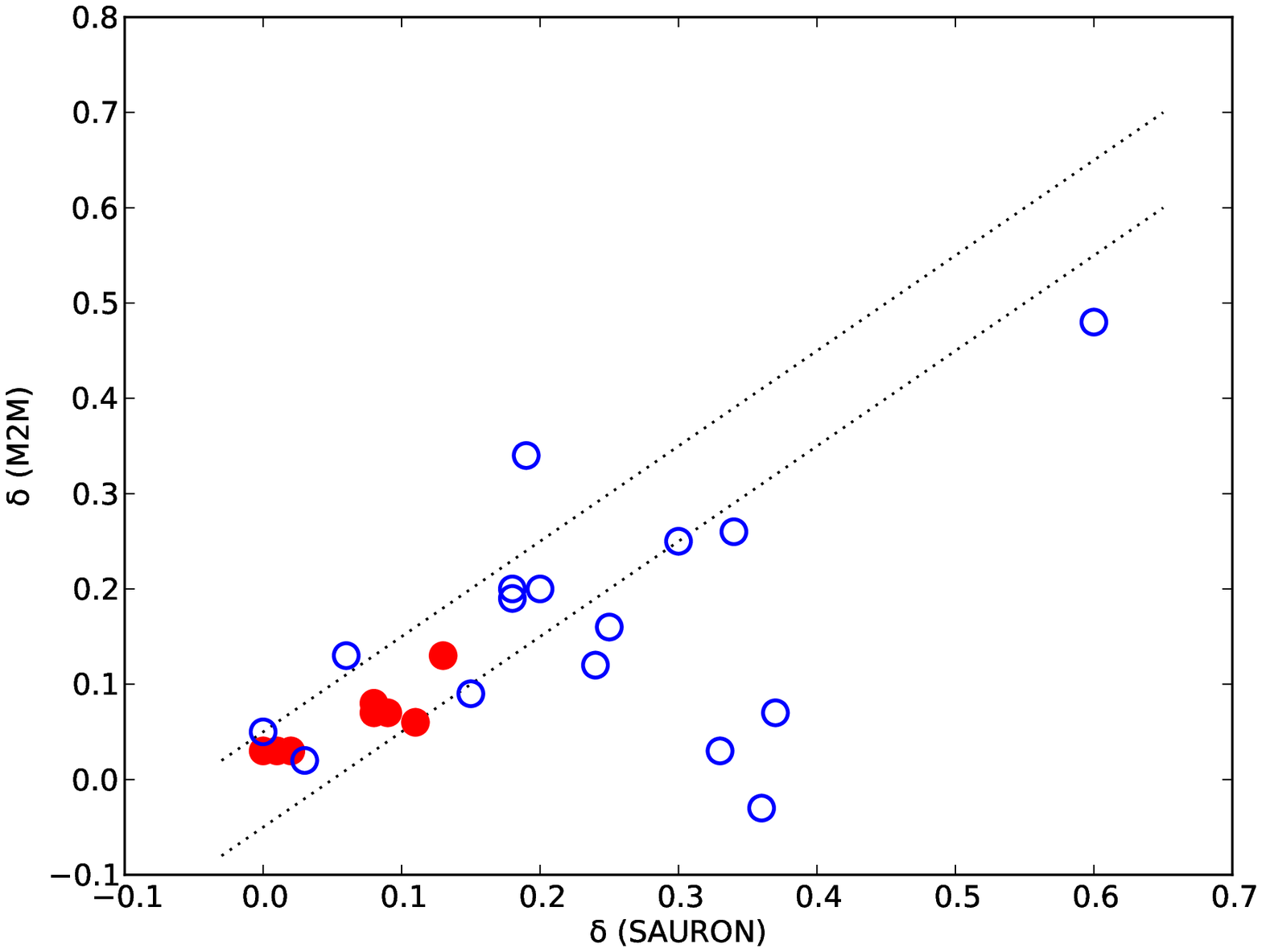}
		\caption[Global anisotropy $\delta $parameter comparison]{Comparison between the M2M and SAURON determined values of the global anisotropy parameter $\delta$. Slow rotating galaxies are plotted with red solid markers. The black dotted lines indicate $\pm 0.05$ of the SAURON values recorded in \citet{SauronX2007}.}
		\label{fig:anicomp1}
\end{figure}
\begin{figure*}
		\centering
		\includegraphics[width=56mm]{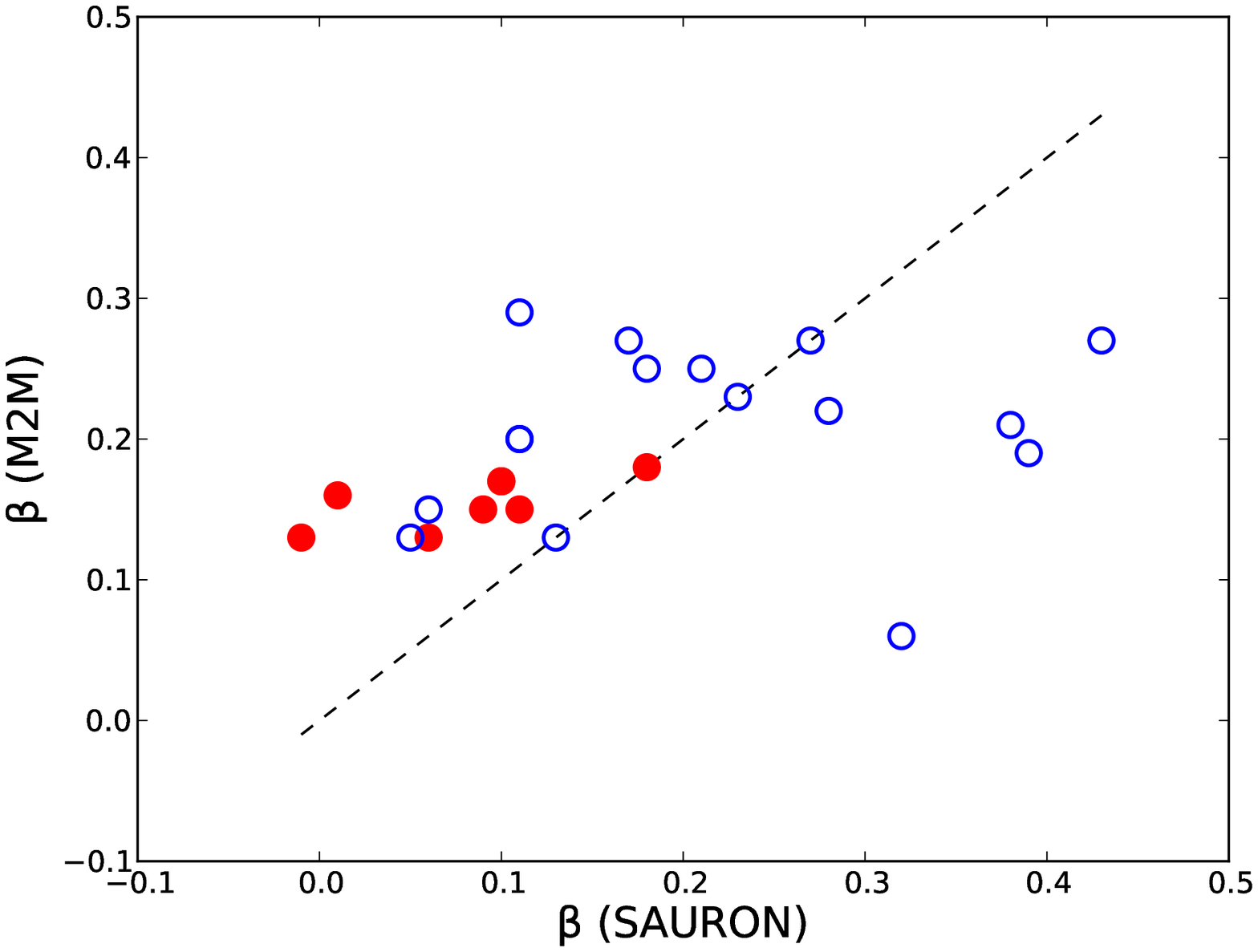}
		\includegraphics[width=56mm]{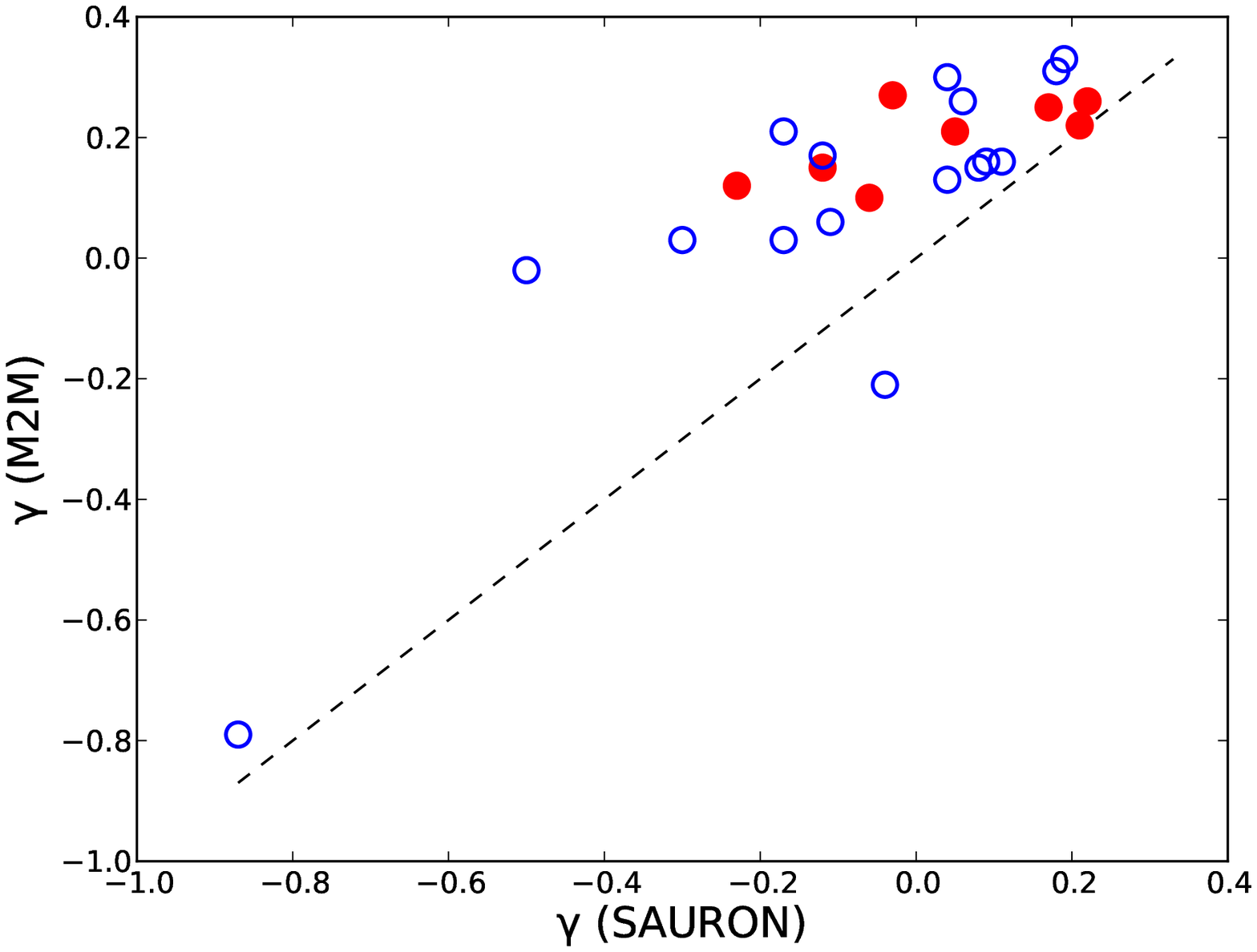}
		\includegraphics[width=56mm]{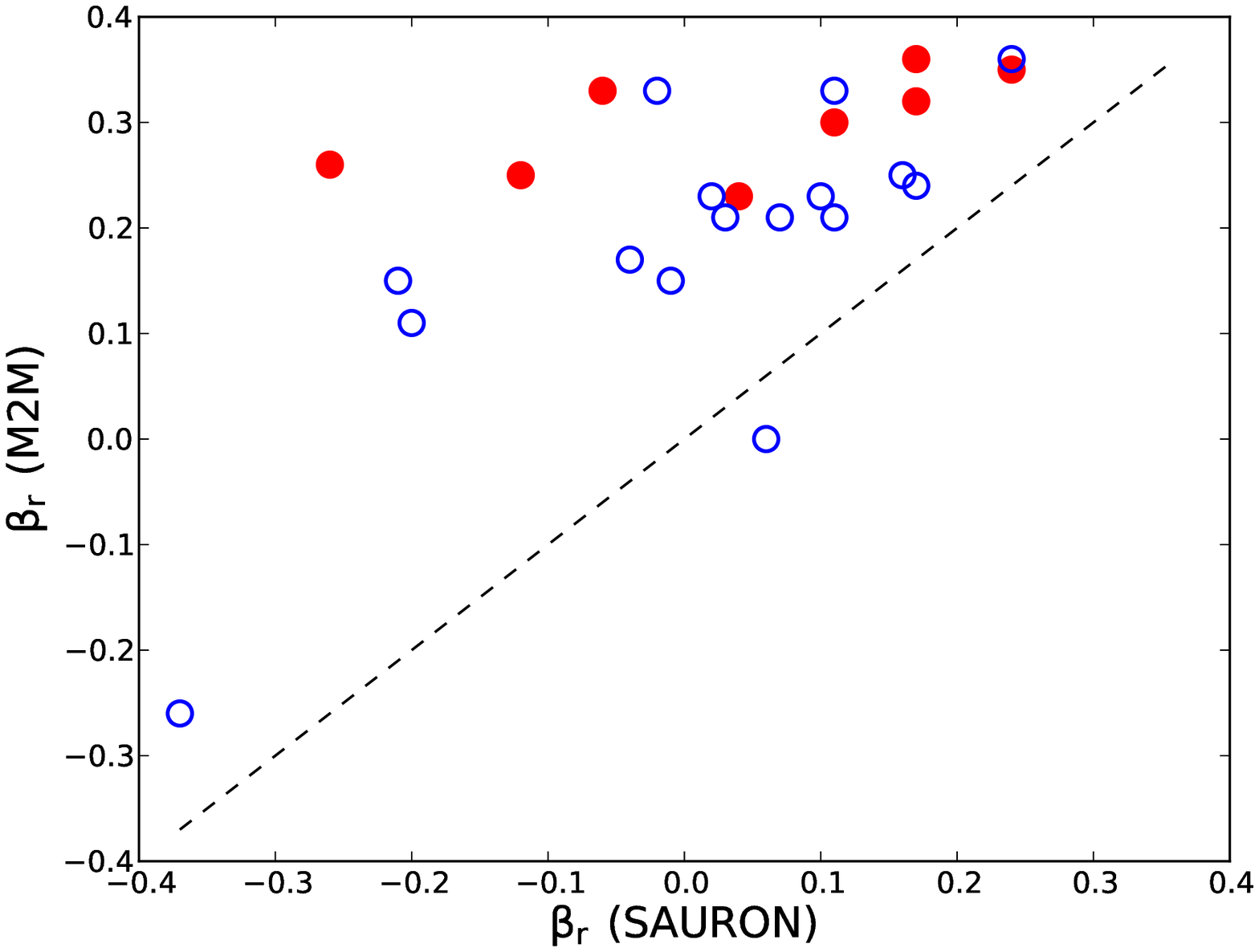}
		\caption[Anisotropy parameters comparison]{Comparison between the M2M and SAURON determined anisotropy parameters $\beta$, $\gamma$ and $\beta _r$. Slow rotating galaxies are plotted with red solid markers. The black dashed lines indicate equality of the M2M and SAURON values.}
		\label{fig:anicomp2x}
\end{figure*}
Figures \ref{fig:anicomp1} and \ref{fig:anicomp2x} compare the M2M parameter values against the SAURON values.  Two main differences can be seen, the first being that in the M2M results the slow rotating galaxies are more clustered in parameter space. The second is that from Figure \ref{fig:aniparams} the number of galaxies exhibiting tangential anisotropy has reduced and we examine this in more detail below.

\begin{table*}
	\centering
	\caption{Anisotropy parameter analysis}
	\label{tab:anianal}
	\begin{tabular}{cccccl}
		\hline
 & \multicolumn{2}{c}{Number of} & \multicolumn{2}{c}{Number of} \\
 & \multicolumn{2}{c}{Positive Values} & \multicolumn{2}{c}{Negative Values} & Galaxy \\
Parameter & Sauron & M2M & Sauron & M2M & NGC numbers \\
		\hline
	Fast Rotators\\
	$\beta _r$ & 10 & 15 & 6 & 1 & 2974, 4150, 4278, 4473, 4621\\
	$\beta$    & 16 & 16 & 0 & 0\\
	$\gamma$   &  8 & 13 & 8 & 3 & 2974, 4150, 4278, 4621, 4660\\
	$\delta$   & 16 & 15 & 0 & 1 & 4150\\
	Slow Rotators\\
	$\beta _r$ &  5 &  8 & 3 & 0 & 3414, 4458, 4552\\
	$\beta$    &  7 &  8 & 1 & 0 & 4458\\
	$\gamma$   &  4 &  8 & 4 & 0 & 3414, 3608, 4458, 4552\\
	$\delta$   &  8 &  8 & 0 & 0\\
		\hline
	\end{tabular}
	
\medskip
The table shows the numbers of galaxies with positive and negative velocity dispersion anisotropy parameter values as a means of assessing any general change of radial or tangential anisotropy regime for the sample of galaxies.  The `galaxy' column indicates those galaxies where the regime differs between the SAURON and M2M models.
\end{table*}

We start by performing a simple count of the number of positive and negative parameter values - the result is recorded in Table \ref{tab:anianal}.  For the fast rotating galaxies, any change of sign of a parameter between the SAURON and M2M values comes from one of 6 galaxies (NGC 2974, 4150, 4278, 4473, 4621, 4660).  Similarly, for slow rotating galaxies, only 4 galaxies are involved (NGC 3414, 3608, 4458, 4552).  From examining the characteristics of the galaxies, there are no obvious groupings which might help explain the differences between the SAURON and M2M results.  For example, the 10 galaxies include both elliptical and lenticular galaxies and galaxies which are inclined to the line-of-sight or edge-on. All the galaxies have M2M mass-to-light ratios which differ from the SAURON values by $< 7.5\%$, and 7 of the galaxies have M2M $\delta$ anisotropy parameter values within $\pm 0.05$ of the SAURON values.  It should be noted that for a given value of the global anisotropy parameter $\delta$, there is a linear relationship between $\beta$ and $\gamma$ so a range of orbital models and velocity dispersion anisotropies is to be expected given the kinematic observations available.  Including luminosity density as a constraint does not alter the tangential anisotropy result.  Note that the data used in \citet{SauronX2007} (see section 2) differ from the SAURON data release though it is not clear that this would explain the differences.

In the absence of further constraints and perhaps more detailed investigations, we conclude that the differences in the anisotropy parameter values are due to differences in the particle / orbit initial conditions and the modelling methods used resulting in different orbital weightings.

\section{Conclusions}\label{sec:conclusions}
We have undertaken the largest M2M exercise to date and re-analysed 24 elliptical and lenticular galaxies previously analysed with Schwarzschild's method.  We have used the M2M method as far as possible as a `black box' - we have developed no computer code specific to any one galaxy. Where there are modelling parameters to be set or tuned, we have adopted the same strategy and process for all galaxies.

Our M2M implementation has been adapted to use observable data available as a Voronoi tessellation and to handle gravitational potentials derived from a multi-Gaussian expansion and deprojection of a galaxy's surface brightness.  We have identified a computer performance issue with our M2M implementation which may affect other users of the method depending on their implementation and network configuration.  For the future, an improved process, preferably computerised, for setting the global rotation of the system of particles is required.

We achieve reasonable agreement ($14$ out of $24$ galaxies) with the SAURON values of the global anisotropy parameter $\delta$ but our overall assessment is that further (theoretical) investigations of the impact of orbit / particle initial conditions and the resultant orbit weights are required before the differences between the SAURON and M2M methods can be fully understood.  In the M2M case, it may prove to better to calculate an exponentially smoothed version of $\delta$ rather than relying on the end of modelling run particle data.

We have demonstrated that, despite differences in the M2M and Schwarzschild modelling methods, in general the methods are delivering similar mass-to-light ratios for a galaxy.  Whether the slight over estimation (M2M) or under estimation (Schwarzschild) is a real effect or not will only be resolved by using a different sample of galaxies.

\section*{Acknowledgements}
The authors gratefully acknowledge the advice, help and support provided by Michele Cappellari, and thank John Magorrian for an enlightening discussion on Gauss-Hermite series.  The final computer runs were performed on the \textit{Laohu} high performance computer cluster of the National Astronomical Observatories, Chinese Academy of Sciences, with earlier runs being performed on the Jodrell Bank Centre for Astrophysics, University of Manchester, \textit{Coma} cluster.

\bibliographystyle{mn2e}
\bibliography{rjlmodel}

\label{lastpage}
\end{document}